\documentclass[letterpaper,twocolumn,english,prb,superscriptaddress,showpacs]{revtex4}
\usepackage[T1]{fontenc}
\usepackage[latin9]{inputenc}
\usepackage{babel}
\usepackage{verbatim}
\usepackage{amsmath}
\usepackage{amssymb}
\usepackage{graphicx}
\usepackage{esint}
\usepackage[unicode=true,pdfusetitle,
 bookmarks=true,bookmarksnumbered=false,bookmarksopen=false,
 breaklinks=false,pdfborder={0 0 1},backref=false,colorlinks=false]
 {hyperref}

\makeatletter

\pdfpageheight\paperheight
\pdfpagewidth\paperwidth

\@ifundefined{textcolor}{}
{%
 \definecolor{BLACK}{gray}{0}
 \definecolor{WHITE}{gray}{1}
 \definecolor{RED}{rgb}{1,0,0}
 \definecolor{GREEN}{rgb}{0,1,0}
 \definecolor{BLUE}{rgb}{0,0,1}
 \definecolor{CYAN}{cmyk}{1,0,0,0}
 \definecolor{MAGENTA}{cmyk}{0,1,0,0}
 \definecolor{YELLOW}{cmyk}{0,0,1,0}
}

\usepackage{babel}

\makeatother

\begin{document}

\title{Magnetic phases in the one-dimensional Kondo chain on a metallic
surface}

\author{Alejandro M. Lobos}

\affiliation{Condensed Matter Theory Center, Department of Physics, University
of Maryland, College Park, Maryland 20742-4111, USA.}

\author{Miguel A. Cazalilla}

\affiliation{Donostia International Physics Center (DIPC), Manuel de Lardiz{\'a}bal
4, 20018 San Sebasti{\'a}n, Spain }

\affiliation{Centro de F{\'i}sica de Materiales (CFM), Centro Mixto CSIC-UPV/EHU,
Edificio Korta, Avenida de Tolosa 72, 20018 San Sebasti{\'a}n, Spain.}

\author{Piotr Chudzinski}

\affiliation{DPMC-MaNEP, University of Geneva, 24 Quai Ernest-Ansermet CH-1211
Geneva, Switzerland.}

\date{\today}
\begin{abstract}
We study the low-temperature properties of a one-dimensional spin-$1/2$
chain of magnetic impurities coupled to a (normal) metal environment
by means of anisotropic Kondo exchange. In the case of easy-plane
anisotropy, we obtain the phase-diagram of this system at $T=0$.
We show that the in-plane Kondo coupling destabilizes the Tomonaga-Luttinger
phase of the spin-chain, and leads to two different phases: i) At
strong Kondo coupling, the spins in the chain form Kondo singlets
and become screened by the metallic environment, and ii) At weak and
intermediate Kondo coupling, we find a novel dissipative phase characterized
by diffusive gapless spin excitations. The two phases are separated
by a quantum critical point of the Wilson-Fisher universality class
with dynamical exponent $z\simeq2$. 
\end{abstract}


\maketitle

\section{\label{sec:intro}Introduction}

Magnetic structures of atomic size provide the smallest solid-state
systems in which it is possible to store (quantum) information.\cite{Wiesendanger09_Spins_on_surfaces_review}
The possibility to build and manipulate such atomic-scale magnetic
structures has been demonstrated in recent experiments using scanning
tunneling microscopy (STM),\cite{Wiesendanger09_Spins_on_surfaces_review,jamneala01,Hirjibehedin06_Mn_spin_chains,Zhou10_RKKY_on_atomic_scale,Serrate10_Imaging_and_manipulating_spin_of_individual_atoms}
a fact that paves the way for the realization of spin-devices of nanoscopic
size.

Besides the interest in applications, magnetic systems at the nanoscale
constitute an excellent playground to address fundamental questions
in condensed matter physics. For instance, magnetic impurities inside
a metallic host have shown clear evidences of the Kondo effect in
the scanning tunneling spectra (STS).~\cite{Li98_Kondo_effect_on_single_adatoms,Madhavan98_Tunneling_into_single_Kondo_adatom,knorr02}
The Kondo effect (i.e., the spin-compensation of a localized magnetic
moment by conduction electrons in a metal) is one of the most paradigmatic
phenomena in many-body physics.\cite{hewson} On the other hand, magnetic
atoms inside a metal can interact non-locally via the electronic medium
through the Ruderman-Kittel-Kasuya-Yosida (RKKY) exchange coupling,\cite{ruderman54}
which is responsible for the magnetic properties of many heavy-fermion
materials\cite{Lohneysen99_Ce_review} and for the giant-magnetoresistance
effect in layered magnetic heterostructures.\cite{Baibich88_GMR_in_layered_magnetic_superlattices}
Direct evidence of RKKY interaction at the atomic scale (i.e., among
pairs of magnetic Fe or Co atoms) has been reported recently in STM
experiments.\cite{Zhou10_RKKY_on_atomic_scale,Wahl07_Exchange_Interaction_between_Single_Magnetic_Adatoms}
Due to the ability to control the distance between magnetic atoms
using the STM tip, the RKKY interaction can be, in principle, tuned
from ferromagnetic to antiferromagnetic, and this oscillating feature
has been clearly revealed in recent experiments showing spin-polarized
STM maps.\cite{Zhou10_RKKY_on_atomic_scale}

By depositing atoms one by one, STM also enables to build magnetic
structures where the dimensionality is gradually changed from the
zero-dimensional (0D) limit to the one-dimensional (1D) case. In particular,
linear arrays of up to 10 magnetic Mn atoms\cite{Hirjibehedin06_Mn_spin_chains}
and, more recently, antiferromagnetic chains made of 8 and more Fe
atoms\cite{Loth12_Bistability_in_atomic_scale_AFM} have been assembled
and analyzed with STM and inelastic electron tunnel spectroscopy (IETS).
From the theoretical point of view, these low-dimensional magnetic
systems are of interest due to the prominent effect of quantum fluctuations,
which inhibit magnetic order and, at low temperatures, lead to quantum
phases with exotic properties.\cite{giamarchi_book_1d} 
\begin{figure}
\begin{centering}
\includegraphics[viewport=0 300 800 600,clip,scale=0.35]{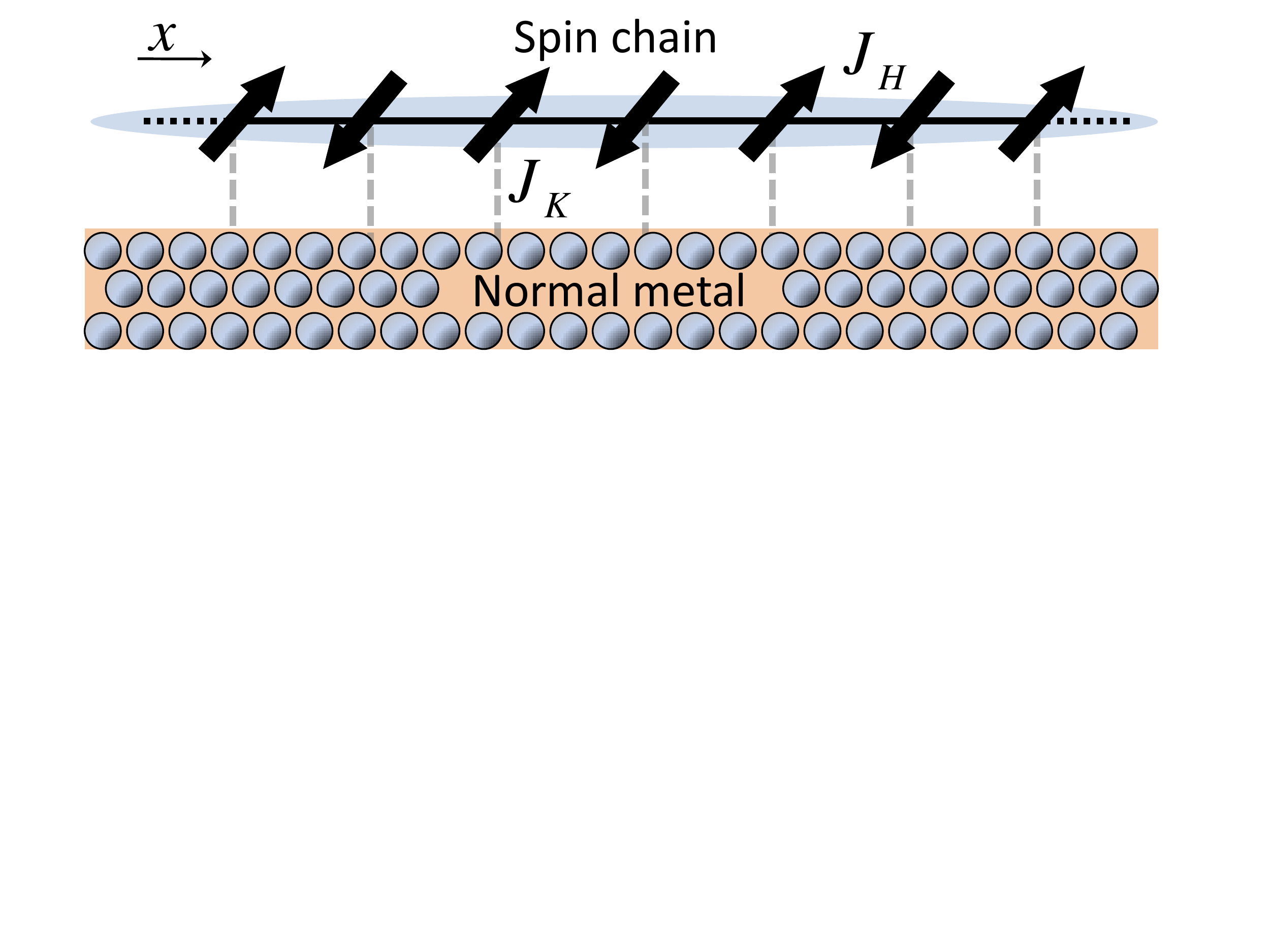} 
\par\end{centering}

\caption{\label{fig:system-1}Scheme of the system representing a $S=1/2$
spin chain coupled to a metallic environment through the Kondo Hamiltonian.
The presence of the metallic surface breaks the rotational and inversion
symmetry, inducing in-plane anisotropy. }
\end{figure}

Motivated by the experimental progress described above, in this work
we study a 1D chain of spin $S=1/2$ magnetic atoms (i.e. Kondo impurities)
embedded in (or, rather, deposited on) a host such like a metallic
surface (cf. Fig.\ref{fig:system-1}). The magnetic atoms are coupled
to each other and to the metallic host by means of an anisotropic
exchange. The anisotropic coupling between the magnetic atoms may
be regarded as a consequence of, e.g., a Dzyaloshinskii-Moriya interaction
resulting from the spin-orbit coupling in the host. From a different
perspective, this system also provides a realization of a 1D dissipative
system, where the interplay between quantum fluctuations and dissipation
can have important consequences for the quantum phase diagram.\cite{caldeira&leggett81}
Examples of other physical realizations of 1D dissipative systems
can be found in e.g. resistively shunted 1D Josephson junctions arrays,\cite{schoen90_review_ultrasmall_tunnel_junctions,fazio01_review_superconducting_networks,Tewari06_Dissipate_locally_couple_globally,goswami06_josephson_array,refael07_SN_transition_in_grains&nanowires,Lobos11_sit_in_dissipative_jjas}
Tomonaga-Luttinger liquids coupled to dissipative baths,\cite{castroneto97_open_luttinger_liquids,cazalilla06_dissipative_transition,artemenko07_longrange_order_in_1d}
superconducting wires coupled to diffusive metals \cite{cazalilla06_dissipative_transition,lobos09_dissipation_scwires,Lobos10_Dissipative_phase_fluctuations}
and 1D ultra-cold atom gases embedded in a Bose-Einstein condensate.~\cite{Orth08_Dissipative_quantum_Ising_model_in_cold_atoms}

Using a host of analytical methods, which include bosonization and
renormalization group methods, we investigate the effects of the metallic
environment on the spin chain. At $T=0$, we predict that the Tomonaga-Luttinger
liquid (TLL) phase of the impurity spin-chain will be destabilized
by the (in-plane) Kondo interaction, $J_{K}^{\perp}$, with the metallic
host. Turning on $J_{K}^{\perp}$, yields a dissipative phase, whose
ground state exhibits long-range order (LRO) of the in-plane magnetization.
Upon increase of $J_{K}^{\perp}$, we predict a quantum phase transition
towards a disordered Kondo-singlet phase, where the spins of 1D chain
are locally screened by the host electrons, and the LRO along the
chain is destroyed.

This work is organized as follows. In Section \ref{sec:model} we
present the model for the anisotropic spin chain coupled to a normal
metal, in Section \ref{sec:results} we show the our main results,
divided into weak-coupling (Sec. \ref{sub:weak_coupling_RG}) and
strong-coupling (\ref{sub:strong_coupling}) treatment of the Kondo
Hamiltonian, in Section \ref{sec:summary} we present a summary and
conclusions, and finally, the details of the calculations are in Appendixes
\ref{sec:RG_free-energy} and \ref{sec:spin_correlation_function}.

\section{Model\label{sec:model}}

The Hamiltonian of the spin-chain system embedded in a metallic host
can be split into three terms: 
\begin{equation}
\mathcal{H}=\mathcal{H}_{C}+\mathcal{H}_{F}+\mathcal{H}_{K},\label{eq:H_total}
\end{equation}
 where $\mathcal{H}_{C}=\sum_{ij,\alpha,\beta}\tilde{J}_{H}^{\alpha\beta}\left(ij\right)\tilde{S}_{i}^{\alpha}\tilde{S}_{j}^{\beta}$,
describes the (static) interactions between the magnetic atoms in
the chain, $\mathcal{H}_{K}$ accounts for the coupling of the chain
to the metallic host, and $\mathcal{H}_{F}$ describes the host electrons.
In a real experimental system the physical origin of the term $\mathcal{H}_{C}$
is not always easy to identify. Although in general the RKKY interaction
is believed to mediate magnetic interactions between magnetic adatoms
deposited on metals (e.g., Co adatoms deposited on the top of Cu(100)
surfaces\cite{Stepanyuk_Magnetic_properties_of_Co_clusters_on_Cu100,Wahl07_Exchange_Interaction_between_Single_Magnetic_Adatoms,Zhou10_RKKY_on_atomic_scale}),
more complicated situations, requiring detailed first-principles calculations
to identify the source of magnetic coupling, might arise. Determining
the physical origin of $\mathcal{H}_{C}$ is however beyond the scope
of the present work, and will not be relevant for our purposes in
what follows. Here we assume quite generally that the coupling $\tilde{J}_{H}^{\alpha\beta}\left(ij\right)$
decays with the distance between impurities, allowing to truncate
the interaction at nearest neighbors. Thus, we consider the following
model 
\begin{align}
\mathcal{H}_{C} & =\sum_{j}\frac{\tilde{J}_{H}^{\perp}}{2}\left(\tilde{S}_{j+1}^{+}\tilde{S}_{j}^{-}+\tilde{S}_{j}^{-}\tilde{S}_{j+1}^{+}\right)+\tilde{J}_{H}^{z}\tilde{S}_{j}^{z}\tilde{S}_{j+1}^{z}\nonumber \\
 & +\sum_{j}D\hat{\mathbf{z}}\cdot\left(\tilde{\mathbf{S}}_{j}\times\tilde{\mathbf{S}}_{j+1}\right).\label{eq:hamc}
\end{align}
 For definiteness, we assume here that the index $j$ runs along the
$\hat{x}-$axis (cf. Fig. \ref{fig:system-1}). The last term in Eq.~\eqref{eq:hamc}
is a Dzyaloshinskii-Moriya (DM) interaction, which results from the
spin-orbit coupling of the electrons at the surface of the metallic
host. The symmetry conditions for the DM interaction to exist are
rarely met in the bulk of typical metals. However, for an impurity
chain that lies on a metallic surface where inversion symmetry is
broken, a DM interaction-term is in principle expected.\cite{Carbone11_Self-assembled_magnetic_networks_on_surfaces}
The above spin-chain model, Eq.~\eqref{eq:hamc}, can be mapped onto
a 1D Heisenberg-Ising (XXZ) $S=1/2$ chain by the following transformation\cite{Garate10_Anisotropic_spin_chains}
\begin{align}
S_{j}^{\pm} & =e^{i\eta j}\:\tilde{S}_{j}^{\pm},\label{eq:transf_Spm}\\
S_{j}^{z} & =\tilde{S}_{j}^{z},\label{eq:transf_Sz}
\end{align}
 where $\eta=\tan^{-1}(D/\tilde{J}_{H})$ ($0\le\eta\le\pi/2$). Thus,
Eq.~\eqref{eq:hamc} becomes 
\begin{align}
\mathcal{H}_{XXZ} & =\sum_{j}\frac{J_{H}^{\perp}}{2}\left(S_{j}^{+}S_{j+1}^{-}+S_{j}^{-}S_{j+1}^{+}\right)+J_{H}^{z}S_{j}^{z}S_{j+1}^{z},\label{eq:H_xxz}
\end{align}
 with $J_{H}^{\perp}=\tilde{J}_{H}^{\perp}/\cos\eta$ and $J_{H}^{z}=\tilde{J}_{H}^{z}$.
Although the DM interaction in metallic surfaces is small\cite{Carbone11_Self-assembled_magnetic_networks_on_surfaces}
(i.e., $D/J_{H}\ll1$), it can dramatically affect the properties
of the spin chain. In particular, if the initial impurity-chain Hamiltonian
(\ref{eq:hamc}) is isotropic (i.e., $\tilde{J}_{H}^{\perp}=\tilde{J}^{z}$),
the above transformation (\ref{eq:transf_Spm}) maps it onto an XXZ
spin chain with easy-plane anisotropy: $\left|J_{H}^{\perp}/J_{H}^{z}\right|=\left(\cos\eta\right)^{-1}>1$.
Therefore, from here on, we shall focus on the case of the XXZ spin
chain with in-plane anisotropy. Under these assumptions, it is well
known that the low-energy sector of Hamiltonian Eq. (\ref{eq:H_xxz})
maps onto the $XY-$model, whose spectrum is described in terms of
gapless spinon modes and exhibits power-law magnetic correlations\cite{affleck_houches,giamarchi_book_1d}
$\left\langle S_{j}^{+}S_{j+n}^{-}\right\rangle \sim\left|n\right|^{-\nu}$.

The coupling between the XXZ chain and the metal is described by the
following anisotropic Kondo exchange interaction\cite{hewson}

\begin{align}
\mathcal{H}_{K} & =\sum_{j}\frac{1}{2}J_{K}^{z}S_{j}^{z}\left[c_{\uparrow}^{\dagger}\left(\mathbf{R}_{j}\right)c_{\uparrow}\left(\mathbf{R}_{j}\right)-c_{\downarrow}^{\dagger}\left(\mathbf{R}_{j}\right)c_{\downarrow}\left(\mathbf{R}_{j}\right)\right]\nonumber \\
 & +\frac{J_{K}^{\perp}}{2}e^{i\eta j}S_{j}^{+}c_{\downarrow}^{\dagger}\left(\mathbf{R}_{j}\right)c_{\uparrow}\left(\mathbf{R}_{j}\right)+\text{H.c.},\label{eq:H_K}
\end{align}
 where Eqs. (\ref{eq:transf_Spm}) and (\ref{eq:transf_Sz}) have
been used. In Eq.(\ref{eq:H_K}) every spin $\mathbf{S}_{j}$ in the
chain is coupled to the fermionic spin-density of the bath. The operator
$c_{\sigma}^{\dagger}\left(\mathbf{R}\right)$ creates an electron
with spin projection $\sigma$ at position $\mathbf{R}=\left(x,y,z\right)$
in the metal, and for a chain site $\mathbf{R}_{j}=\left(ja_{0},0,0\right)$
with $a_{0}$ the lattice parameter of the impurity chain. Here we
also assume an anisotropic Kondo interaction with in-plane anisotropy
$\left|J_{K}^{\perp}/J_{K}^{z}\right|>1$.

Finally, the dynamics of the electrons in the metallic host is described
in terms of Landau quasi-particles 
\begin{align}
\mathcal{H}_{F} & =\sum_{\mathbf{k},\sigma}\epsilon_{\mathbf{k}}c_{\sigma}^{\dagger}\left(\mathbf{k}\right)c_{\sigma}\left(\mathbf{k}\right)+\cdots,\label{eq:H_F}
\end{align}
where $\epsilon_{\mathbf{k}}$ is the electron dispersion and $c_{\sigma}\left(\mathbf{k}\right)\equiv\int d^{3}\mathbf{R}\; e^{i\mathbf{k.R}}c_{\sigma}\left(\mathbf{R}\right)$
is the Fourier transform of the annihilation operator $c_{\sigma}\left(\mathbf{R}\right)$.
The dots in Eq.~\eqref{eq:H_F} stand for additional terms, such
as spin-obit interactions, whose form needs not be specified. An important
parameter describing the properties of the metallic host is the Fermi
wavevector $k_{F}$. In a real experimental situation, although the
magnetic nanostructure is built on the top of the 2D metallic surface,
$k_{F}$ might have a 3D character due to a non-vanishing overlap
with the bulk conduction states in the metal.\cite{knorr02}

Intuitively, the Hamiltonian Eq.\eqref{eq:H_total} encodes a competition
between the Heisenberg interaction, which favors correlations along
the spin chain, and the Kondo coupling, which tends to screen locally
the impurity-spins and promotes a non-magnetic ground state. Our description
is therefore very similar to the well-known case of heavy-fermion
materials described by the (3D) Kondo lattice model, where this competition
between the Kondo and RKKY interactions is believed to be responsible
for the unusual quantum critical properties and complex magnetic phase
diagram.\cite{hewson,CastroNeto00_NFL_in_U_and_Ce_alloys,Si10_Heavy_fermions_and_phase_transitions}
In that context, it has been shown (See e.g., Ref. \onlinecite{CastroNeto00_NFL_in_U_and_Ce_alloys})
that although both interactions are due to coupling of spins with
the same Fermi sea, they can be treated separately since the RKKY
interaction originates from electronic states deep inside the Fermi
sea, while the Kondo effect is purely a Fermi-surface effect. In our
particular case the interplay between Heisenberg and Kondo interaction
is quite subtle due to the additional effects of plane-anisotropy
and reduced dimensionality of the spin-chain, leading to a counterintuitive
cooperation effect in a certain regime of parameters and to a non-trivial
phase diagram at $T=0$ (cf. Fig. \ref{fig:phase_diagram} below). 

One should note that our model significantly differs from previous
approaches to the 1D Heisenberg-Kondo mode (1DHKM)~ \cite{zachar_exotic_kondo,Zachar01_Staggered_phases_1D_Kondo_Heisenberg_model,Strong94_Competition_between_Heisenberg_and_Kondo_in_1D}
or the 1D Kondo-lattice model (1DKLM).\cite{zachar_kondo_chain_toulouse,Sikkema97_Spin_gap_in_a_doped_Kondo_chain,tsunetsugu_kondo_1d,Novais02_Phase_diagram_of_the_anisotropic_Kondo_chain,Braunecker09_Nuclear_magnetism_in_C13_nanotubes}
Those works assumed entirely 1D situations in which a 1D electron
gas acting as a host is coupled either with a 1D spin chain (1DHKM)
or with a set of independent (0D) spins (1DKLM), a $J_{H}=0$ limit
of the 1DHKM. In those works the 1D character of the bath allowed
not only for a straightforward application of 1D methods (e.g. bosonization
or DMRG), but it also contained hidden constrains like spin-charge
separation or a reduction of the Fermi surface only to two points
(thus allowing for a very limited number of scattering channels).
In particular, it was shown that the magnetic scattering on spin impurities
opens a spin gap in the spectrum of the 1D fermionic bath.\cite{zachar_kondo_chain_toulouse,Sikkema97_Spin_gap_in_a_doped_Kondo_chain,Zachar01_Staggered_phases_1D_Kondo_Heisenberg_model}
This result is in clear contrast with our assumption in Eq. (\ref{eq:H_F}),
where we take the higher-dimensional bath to be unaffected by low
dimensional impurities. The principal aim of those previous works
was also different, i.e. to establish analogies with phases present
in heavy-fermion compounds, as clearly expressed in a review article.\cite{tsunetsugu_kondo_1d}

In contrast, our model is closer to the experimental situations described
in Sec. \ref{sec:intro}, where the spin chain is embedded in or lies
on a higher-dimensional metallic host with a large Fermi surface and
large Fermi energy. Consequently, in our work many of the above-mentioned
contraints existing in the entirely 1D geometry are released. The
reason why this drastically changes the physics of the problem results
from the absence of the Nozi{\`e}res' ``exhaustion'' problem\cite{nozieres74}:
in our case, a higher-dimensional bath ensures the presence of enough
conduction electrons to screen the magnetic impurities on the scale
$\sim k_{F}^{-1}$ (cf. Sec. \ref{sub:strong_coupling} below for
details). One should keep in mind that taking the higher-dimensional
bath is at the core of this paper and is in fact crucial to justify
the local approximation in Sec. \ref{sub:strong_coupling} (i.e.,
independent local-bath approximation).

In the following, we explore the quantum critical properties of model
Eq. (\ref{eq:H_total}) at low-energies and for different regimes
of parameters $J_{H}^{\perp},J_{H}^{z},J_{K}^{\perp}$ and $J_{K}^{z}$.
Throughout we shall use units where $\hbar=1$.

\section{\label{sec:results}Results}

\subsection{Weak coupling scaling analysis\label{sub:weak_coupling_RG}}

In the regime where the Heisenberg coupling $J_{H}^{\perp}$ dominates
(i.e., $J_{H}^{\perp}$ is much larger than $J_{K}^{\perp}$ and $J_{K}^{z}$),
a good starting point is to treat the Kondo coupling, $\mathcal{H}_{K}$
{[}cf. Eq.~(\ref{eq:H_K}){]}, as a small perturbation to the isolated
spin-chain Hamiltonian $\mathcal{H}_{XXZ}$ {[}cf. Eq.~(\ref{eq:H_xxz}){]}.
In this case, Hamiltonian Eq. (\ref{eq:H_xxz}) can be studied within
the framework of Abelian bosonization, \cite{giamarchi_book_1d} which
allows to map it onto the continuous XY Hamiltonian 
\begin{align}
\mathcal{H}_{XXZ} & =\frac{1}{2\pi}\int dx\;\left[\frac{u}{K}\left(\nabla\Phi\right)^{2}+uK\left(\nabla\Theta\right)^{2}\right]+\dots\label{eq:H_XXZ_Bosonized}
\end{align}
 Here $\Theta\left(x\right),\Phi\left(x\right)$ are conjugate canonical
fields obeying the usual commutation relations $\left[\Theta\left(x\right),\nabla\Phi\left(x^{\prime}\right)\right]=i\pi\delta\left(x-x^{\prime}\right)$.
These fields are continuous in the scale of $a_{0}$, the original
lattice spacing in the chain, and are related to the original spin
operators by\cite{giamarchi_book_1d} 
\begin{align}
S_{j}^{\pm} & =a_{0}S^{\pm}\left(x_{j}\right)=\frac{e^{\mp i\Theta\left(x_{j}\right)}}{\sqrt{2\pi}}\left[e^{ix_{j}\pi/a_{0}}+\cos2\Phi\left(x_{j}\right)\right],\label{eq:S_pm_Bosonized}\\
S_{j}^{z} & =a_{0}S^{z}\left(x_{j}\right)=-\frac{a_{0}}{\pi}\nabla\Phi\left(x_{j}\right)+\frac{e^{ix_{j}\pi/a_{0}}}{\sqrt{\pi}}\cos2\Phi\left(x_{j}\right),\label{eq:S_z_Bosonized}
\end{align}
 where $x_{j}=ja_{0}$ is the position of the $j$-th spin. The model
(\ref{eq:H_XXZ_Bosonized}) describes 1D gapless spinon excitations
in the transverse direction propagating with velocity $u$, and is
parametrized by the dimensionless Luttinger parameter,~\cite{takahashi73,luther_chaine_xxz}
$K=\left[\frac{2}{\pi}\arccos\left(-J_{H}^{z}/J_{H}^{\perp}\right)\right]^{-1}$,
which determines the decay of the correlation functions in the chain,
e.g., $\left\langle S^{+}\left(x\right)S^{-}\left(0\right)\right\rangle \sim\left|x\right|^{-1/2K}$.
The isotropic Heisenberg model is recovered for the particular value
$K=1/2$. As discussed above, in this work we focus on the regime
of easy-plane anisotropy, which corresponds to the condition that
$K>1/2$. The ellipsis in \eqref{eq:H_XXZ_Bosonized} stands for additional
perturbations, such as the sine-Gordon term $\sim\cos4\Phi\left(x\right)$,
which are irrelevant in the renormalization-group (RG) sense for $K>\frac{1}{2}$
and will be neglected.

The continuum limit of the Kondo Hamiltonian, Eq.~\eqref{eq:H_K},
reads 
\begin{align}
\mathcal{H}_{K}= & \int dx\;\frac{J_{K}^{z}}{k_{F}^{3}}S^{z}\left(x\right)s^{z}\left(x\right)\nonumber \\
 & +\int dx\;\frac{J_{K}^{\perp}}{2k_{F}^{3}}\left[e^{iq_{DM}x}S^{+}\left(x\right)s^{-}\left(x\right)+\text{H.c.}\right],\label{eq:H_K_Bosonized}
\end{align}
 where we have defined wavevector $q_{DM}\equiv\eta/a_{0}$ associated
to the DM interaction, and introduced the factors of the Fermi momentum
$k_{F}$ in order for the Kondo couplings to have dimensions of energy.
We have also defined the spin-density operator of the host electrons
as 
\begin{align}
s^{a}\left(\mathbf{R}\right) & \equiv\sum_{\sigma,\sigma^{\prime}}c_{\sigma}^{\dagger}\left(\mathbf{R}\right)\left[\frac{\boldsymbol{\sigma}^{a}}{2}\right]_{\sigma,\sigma^{\prime}}c_{\sigma^{\prime}}\left(\mathbf{R}\right),
\end{align}
 with $\boldsymbol{\sigma}^{a}$ $\left(a=x,y,z\right)$ the Pauli
matrices. From Eqs. (\ref{eq:S_z_Bosonized}) and (\ref{eq:H_K_Bosonized})
we note that the operator $\nabla\Phi\left(x\right)$ couples to the
uniform component of the spin-density in the electron-gas $s_{u}^{z}\left(x\right)\equiv s^{z}\left(x\right)$,
and the operator $\cos2\Phi\left(x\right)$ couples to the staggered
component $s_{s}^{z}\left(x\right)=e^{ix\pi/a_{0}}s^{z}\left(x\right)$.
On the other hand, the operator $e^{-i\Theta\left(x_{j}\right)}$
couples to the staggered magnetization $s_{s}^{-}\left(x\right)=e^{ix\pi/a_{0}}s^{-}\left(x\right)$.
These contributions to Eq. (\ref{eq:H_K_Bosonized}) have different
scaling dimensions, and we treat them independently in the following
analysis.

Next, we assess the stability of the TLL phase, which is described
by Hamiltonian in Eq.~\eqref{eq:H_XXZ_Bosonized}. To this end, we
consider the leading order corrections to the free energy per unit
of length in the impurity spin-chain. The technical details of this
calculation can be found in Appendix~\ref{sec:RG_free-energy}. We
shall not pursue the stability analysis beyond the leading order in
this work, as our focus here is on the phase diagram in the $K>1/2$
(i.e. $J_{H}^{\perp}>J_{H}^{z}$) regime, for which, as the following
discussion demonstrates, there is only one relevant Kondo coupling,
namely $J_{K}^{\perp}$. A more complete analysis will be reported
elsewhere.~\cite{Lobos12_unpublished_Kondo_Heisenberg} To leading
order in the Kondo couplings, for temperatures $T\ll J_{H}^{\perp}$,
we find 
\begin{align}
\frac{\Delta F}{L} & =g_{z,u}^{2}\frac{KA_{u}^{z}}{2^{6}\pi^{2}u^{2}k_{F}}T^{3}\label{eq:free_nrg1}\\
 & +g_{z,s}^{2}\frac{A_{s}^{z}}{2^{4}k_{F}a_{0}^{2}\pi^{3}}\left(\frac{\pi a_{0}}{u}\right)^{2K}T^{1+2K}\label{eq:free_nrg2}\\
 & +g_{\perp,s}^{2}\frac{A_{s}^{\perp}}{2^{5}k_{F}a_{0}^{2}\pi^{3}}\left(\frac{\pi a_{0}}{u}\right)^{1/2K}T^{1+1/2K},\label{eq:free_nrg3}
\end{align}
 where $L\to\infty$ is the impurity chain length and the dimensionless
couplings $g_{z,u}\equiv J_{K}^{z}/v_{F}k_{F}$, $g_{z,s}\equiv J_{K}^{z}/v_{F}k_{F}$
and $g_{\perp,s}\equiv J_{K}^{\perp}/v_{F}k_{F}$ have been introduced;
$A_{u}^{z}$ , $A_{s}^{z}$ and $A_{s}^{\perp}$ are non-universal
numerical coefficients. As explained in the Appendix~\ref{sec:RG_free-energy},
Eqs. (\ref{eq:free_nrg1})-(\ref{eq:free_nrg3}) reflect the fact
that the metallic host exhibits an Ohmic spectrum of magnetic excitations
over a broad range of momentum transfer along the spin-chain direction.
This means that the metallic host contributes to $\Delta F$ effectively
as if it was a local fermionic bath, i.e., the spin-spin correlation
function, $\chi^{ab}\left(\mathbf{R}_{1}-\mathbf{R}_{2},\tau_{1}-\tau_{2}\right)\equiv-\left\langle T_{\tau}s^{a}\left(\mathbf{R}_{1},\tau_{1}\right)s^{b}\left(\mathbf{R}_{2},\tau_{2}\right)\right\rangle _{0}$,
behaves effectively as a `local' function $\chi^{ab}\left(\mathbf{R}_{1}-\mathbf{R}_{2},\tau_{1}-\tau_{2}\right)\propto\delta_{\mathbf{R}_{1},\mathbf{R}_{2}}/\left(\tau_{1}-\tau_{2}\right)^{2}$
(cf. Sec. \ref{sec:RG_free-energy}). We will come back to this point
in the next Sec. \ref{sub:strong_coupling}.\textbf{ }

The stability of the TLL phase with respect to the perturbation $\mathcal{H}_{K}$
can be now assessed by comparing the scaling with temperature of $\Delta F$
and the free energy of the spin-chain chain, $F_{0}$, when decoupled
from the environment, which behaves as~\cite{giamarchi_book_1d,gogolin_book}
$F_{0}\sim T^{2}$. Thus we look for divergences in the perturbative
corrections to $\Delta F/F_{0}$ as the temperature is gradually decreased
towards the ground state (i.e. $T=0$). From~\eqref{eq:free_nrg3},
it can be seen that the term $\propto g_{\perp,s}^{2}$ yields a divergent
contribution to $\Delta F/F_{0}$, which signals an instability of
the TLL phase.

To make contact with the renormalization group (RG), we shall define
the effective couplings $g_{z,u}\left(\ell\right)\equiv g_{z,u}e^{-\left(1/2\right)\ell}$,
$g_{z,s}\left(\ell\right)\equiv g_{z,s}e^{\left(1/2-K\right)\ell}$
and $g_{\perp,s}\left(\ell\right)\equiv g_{\perp,s}e^{\left(1/2-1/4K\right)\ell}$,
where $\ell\equiv\ln\left(\Lambda_{0}/T\right)$ and $\Lambda_{0}\sim J_{H}^{\perp}$
is the high-energy cutoff for the effective low-energy description
of the spin-chain in terms of Eq.~\eqref{eq:H_XXZ_Bosonized}. Decreasing
the temperature a bit towards the ground state (i.e. $T=0$) can regarded
as an infinitesimal change of $\ell\rightarrow\ell+\delta\ell$, and
the corresponding change (``flow'') of the effective couplings can
described by the following set of differential equations 
\begin{align}
\frac{dg_{z,u}\left(\ell\right)}{d\ell} & =-\frac{1}{2}g_{z,u}\left(\ell\right),\label{eq:RG_g_zu}\\
\frac{dg_{z,s}\left(\ell\right)}{d\ell} & =\frac{1}{2}\left(1-2K\right)g_{z,s}\left(\ell\right),\label{eq:RG_g_zs}\\
\frac{dg_{\perp,s}\left(\ell\right)}{d\ell} & =\frac{1}{2}\left(1-\frac{1}{2K}\right)g_{\perp,s}\left(\ell\right).\label{eq:RG_g_perps}
\end{align}
 Alternatively, we can regard these equations as describing the change
in effective (dimensionless) couplings of an equivalent (coarse-grained)
system with a reduced high-energy cutoff $\Lambda(\ell)=e^{-\ell}\Lambda_{0}$.
This interpretation means that for in-plane anisotropy where $K>1/2$,
the couplings $g_{z,u}\left(\ell\right)$ and $g_{z,s}\left(\ell\right)$
decrease as the system is coarse-grained by integrating out the high-energy
degrees of freedom and become irrelevant (in the RG sense). In other
words, the terms in $\mathcal{H}_{K}$ proportional to those couplings
yield subleading corrections and therefore can be neglected as $T\to0$.
On the other hand, $g_{\perp,s}\left(\ell\right)$ is a relevant (in
the RG sense) perturbation, which, as $T\to0$ yields an dominant
correction and destabilizes the TLL phase of the spin-chain described
by Eq.~\eqref{eq:H_XXZ_Bosonized}. Note that both $g_{z,s}\left(\ell\right)$
and $g_{\perp,s}\left(\ell\right)$ have the same scaling dimension
at the Heisenberg point ($K=1/2$), where they are marginally relevant,
and a higher order perturbative analysis is required to fully assess
the stability of the TLL phase.~\cite{Lobos12_unpublished_Kondo_Heisenberg}

Note that the physics described by Eqs. (\ref{eq:RG_g_zu})-(\ref{eq:RG_g_perps})
can be mimicked by an infinite set of fermionic baths, each bath being
locally coupled to only one impurity spin (cf. Fig.~\ref{fig:system}),
which yields a local (i.e. momentum independent) Ohmic spin response:
e.g. $\chi^{+-}(x,\tau),\chi^{zz}(x,\tau)\sim\delta(x)/\tau^{2}$.
As it will be discussed in the next section, this model allows us
to treat the relevant Kondo coupling $J_{K}^{\perp}$ in a non-perturbative
way. In particular, it captures the important (non-perturbative) feature
that the magnetic moment of the impurities will be fully screened
by the metallic environment at large $J_{K}^{\perp}\gg J_{H}^{\perp}$.
In the above analysis, the need for a non-perturbative treatment is
evidenced by the fact that even an infinitesimal value of $J_{K}^{\perp}$
will destabilize the TLL phase for in-plane anisotropy ($K>1/2$).
However, different from the single-impurity Kondo problem, we will
see below that the RG flow does not proceed from the TLL phase into
a strong coupling Kondo-screened phase in a straightforward manner,
but rather, another phase of dissipative nature intervenes between
the TLL and the Kondo phase.

\subsection{\label{sub:strong_coupling}Strong coupling analysis }

\subsubsection{Derivation of an effective 1D model}

As mentioned before, when \textit{$J_{K}^{\perp}$} flows to strong
coupling, the perturbative RG approach used in the previous Section
\ref{sub:weak_coupling_RG} is no longer valid, and we need to study
the physical properties of the spin chain in a different way. The
approach used in this section is motivated by the following arguments:
i) the analysis made in the previous section and in the Appendix \ref{sec:RG_free-energy}
indicates that the most relevant coupling of the spin-chain to the
metal arises from the local sector of the spin-response in the metal
and, ii) at strong-coupling, for a 2D or 3D host, the interference
of two Kondo screening-clouds belonging to spins located at a distance
$a_{0}$ decays rapidly when $a_{0}$ is of the order of a few Fermi
wavelengths (i.e. $\sim k_{F}^{-1}$).~\cite{Andreani93_Two-impurity_Anderson_model_variational,Barzykin00_Kondo_cloud,Simonin07_Kondo_cloud}
In our case, the higher dimensionality of the metallic host as compared
to the spin-chain allows to rule out the well-known exhaustion problem\cite{nozieres_exhaustion}
and the spins in the chain can be considered as independently screened
in the regime $k_{F}a_{0}\gtrsim1$. Experimentally, this is confirmed
by the behavior of the STS Fano line shapes in magnetic Co atoms deposited
on Cu$(100)$ and separated by distances $a_{0}>8\textrm{\AA}$, which
are identical to the single-impurity STS line shapes.\cite{Wahl07_Exchange_Interaction_between_Single_Magnetic_Adatoms}.
This an important difference with respect to the strictly 1D Kondo-lattice
model, where the single Kondo-impurity limit is reached only at distances
$a_{0}\gg\xi_{K}\sim v_{F}/T_{K}$.\cite{zachar_kondo_chain_toulouse}
Since the Kondo temperature is an exponentially small energy scale,
in the purely 1D geometry the single-impurity regime is only reached
at extremely dilute impurity-spin concentrations.\cite{Barzykin00_Kondo_cloud,Simonin07_Kondo_cloud}

\begin{figure}
\begin{centering}
\includegraphics[viewport=0 100 800 300,clip,scale=0.35]{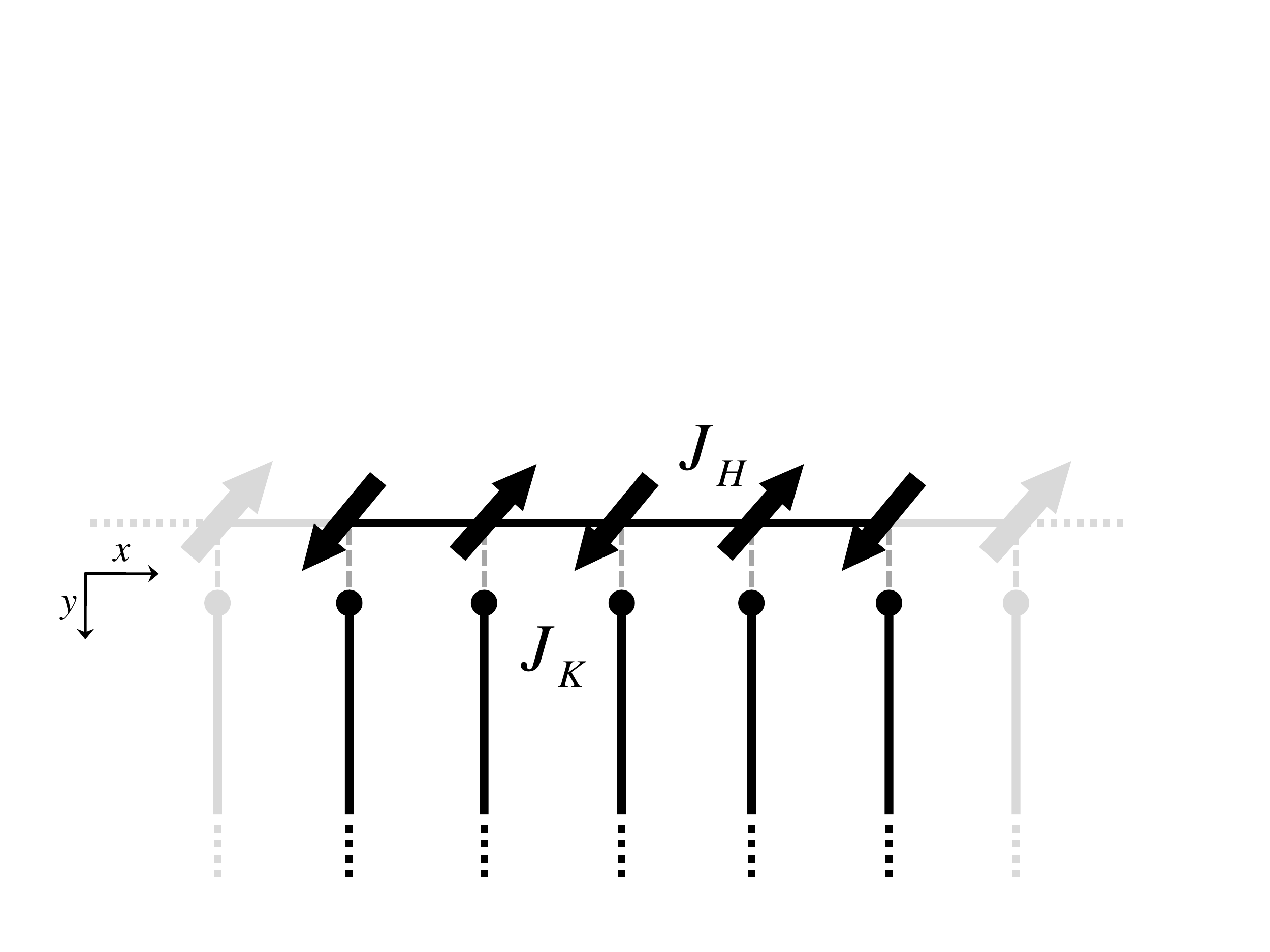} 
\par\end{centering}

\caption{\label{fig:system}Schematic diagram of the spin-chain coupled to
the metallic host, after the independent-bath approximation. Due to
the locality of the dynamic spin-susceptibility, which decays on distances
of the order of $\sim k_{F}^{-1}$, the interference of Kondo-screening
clouds is negligible in the limit $k_{F}a_{0}\gg1$.}
\end{figure}

In our model, the larger dimensionality of the metallic host suggests
that, in the regime where $k_{F}a_{0}\gtrsim1$, it is reasonable
to approximate the Hamiltonian $\mathcal{H}_{F}$ in Eq. (\ref{eq:H_F})
by set of independent fermionic baths (i.e., semi-infinite 1D chains,
cf. Fig. \ref{fig:system}) coupled to each spin in the chain $\mathbf{S}_{i}$,
i.e. 
\begin{align}
\mathcal{H}_{F} & \simeq-t\sum_{ij\sigma}c_{i,j,\sigma}^{\dagger}c_{i,j+1,\sigma}+\text{H.c.},\label{eq:H_F_independent_bath}
\end{align}
 where, the index $i\left(j\right)$ runs along the $\hat{x}\left(\hat{y}\right)$-axis.
 This is a minimal model that captures the competition between Kondo
and Heisenberg interactions. Note that this approximation is consistent
with the local limit of Sec. \ref{sub:weak_coupling_RG}.

In the regime $k_{F}a_{0}\ll1$, an electron at the Fermi energy cannot
distinguish between neighboring individual spins.\cite{Andreani93_Two-impurity_Anderson_model_variational,Simonin06_Two_Anderson_impurities_in_the_Kondo_limit}
Therefore, in that case assuming a local fermionic bath for each impurity
may appear to be a rather uncontrolled approximation. However, whether
this approximation breaks down or not in that regime will be the subject
of future research.\cite{Lobos12_unpublished_Kondo_Heisenberg} Actually,
we may regard the model resulting form Eq.~(\ref{eq:H_F_independent_bath})
as a good starting point for such further investigations.

The advantage of the independent-bath approximation Eq. (\ref{eq:H_F_independent_bath})
is that it allows to use powerful analytical methods which have been
applied successfully to describe the single Kondo-impurity problem.
In the following, we implement the Abelian bosonization approach to
the Kondo problem.\cite{giamarchi_book_1d,gogolin_book,schlottmann_transformation_kondo,emery_kivelson_kondo_review,Kotliar96_Toulouse_points_in_the_generalized_Anderson_model}
To avoid confusion with the previous Sec. \ref{sub:weak_coupling_RG},
note that here bosonization is implemented to describe the \textit{fermionic
1D chains,} and not the spin chain. At low energies the Hamiltonians
$\mathcal{H}_{F}$ and $\mathcal{H}_{K}$ become in the bosonic representation\cite{Kotliar96_Toulouse_points_in_the_generalized_Anderson_model}%
{} 
\begin{align}
\mathcal{H}_{F} & =\sum_{i,\nu=\left\{ c,s\right\} }\frac{v_{F}}{4\pi}\int_{-\infty}^{\infty}dy\;\left(\nabla\phi_{i,\nu}^{R}\left(y\right)\right)^{2},\label{eq:H_F_bosonized}\\
\mathcal{H}_{K} & =\sum_{i}-\frac{2\delta_{s}}{\pi\rho_{0}}S_{i}^{z}\frac{\nabla\phi_{i,s}^{R}\left(0\right)}{\sqrt{2}\pi}\nonumber \\
 & +\frac{J_{K}^{\perp}b_{0}}{2}\left[S_{i}^{+}\frac{e^{-i\sqrt{2}\phi_{i,s}^{R}\left(0\right)}}{2\pi b_{0}}+S_{i}^{-}\frac{e^{i\sqrt{2}\phi_{i,s}^{R}\left(0\right)}}{2\pi b_{0}}\right],\label{eq:H_K_bosonized}
\end{align}
 where the chiral fields $\phi_{i,c}^{R}\left(y\right),\phi_{i,s}^{R}\left(y\right)$
obey the commutation relations $\left[\phi_{i,\nu}^{R}\left(y\right),\phi_{j,\eta}^{R}\left(y^{\prime}\right)\right]=i\pi\text{sign}\left(y-y^{\prime}\right)\delta_{i,j}\delta_{\nu,\eta}$,
and are related to charge and spin density-fluctuations through the
relations $\rho_{i}\left(y\right)=-\frac{1}{\pi}\nabla\phi_{i,c}^{R}\left(y\right)$
and $s_{i}\left(y\right)=-\frac{1}{\pi}\nabla\phi_{i,s}^{R}\left(y\right)$,
respectively.\cite{giamarchi_book_1d} In Eq. (\ref{eq:H_F_bosonized})
$v_{F}$ is the Fermi velocity, and in Eq. (\ref{eq:H_K_bosonized})
$\delta_{s}=\tan^{-1}\left(\pi\rho_{0}J_{K}^{z}b_{0}/4\right)$ is
the scattering phase-shift associated with the potential $J_{K}^{z}S_{i}^{z}/2$,
$\rho_{0}=\left(2\pi v_{F}\right)^{-1}$ the conduction electron density
of states at the Fermi energy, and $b_{0}$ the lattice parameter
in the fermionic chain. For simplicity we assume these parameters
to be identical for all chains.%
{} We then introduce the (Emery-Kivelson) unitary transformation\cite{emery_kivelson_kondo_review}

\begin{align}
\mathcal{U} & =\exp\left[-i\gamma\sum_{i}S_{i}^{z}\phi_{i,s}^{R}\left(0\right)\right].\label{eq:transf_U}
\end{align}
 under which the bosonic field $\nabla\phi_{i,s}^{R}\left(y\right)$
and the spin operator $S_{i}^{+}$ transform as 
\begin{align}
\mathcal{U}^{\dagger}\nabla\phi_{i,s}^{R}\left(y\right)\mathcal{U} & =\left[\nabla\phi_{i,s}^{R}\left(y\right)+\delta\left(y\right)2\pi\gamma S_{i}^{z}\right],\label{eq:transformation_1}\\
\mathcal{U}^{\dagger}S_{i}^{+}\mathcal{U} & =S_{i}^{+}e^{i\gamma\phi_{i,s}^{R}\left(0\right)}.\label{eq:transformation_2}
\end{align}
 Upon this transformation, the model Hamiltonian, Eq.~\eqref{eq:H_total}
transforms as $\tilde{\mathcal{H}}=\mathcal{U}^{\dagger}\mathcal{H}\mathcal{U}=\tilde{\mathcal{H}}_{F}+\tilde{\mathcal{H}}_{K}+\tilde{\mathcal{H}}_{XXZ}$,
with 
\begin{align}
\tilde{\mathcal{H}}_{F} & =\mathcal{H}_{F},\label{eq:H_F_transformed}\\
\tilde{\mathcal{H}}_{K} & =\sum_{i}-\frac{2\tilde{\delta}_{s}}{\pi\rho_{0}}S_{i}^{z}\frac{\nabla\phi_{i,s}^{R}\left(0\right)}{\sqrt{2}\pi}\nonumber \\
 & +\sum_{i}\frac{J_{K}^{\perp}b_{0}}{2}\left[S_{i}^{+}\frac{e^{-i\left(\sqrt{2}-\gamma\right)\phi_{i,s}^{R}\left(0\right)}}{2\pi b_{0}}+\text{H.c.}\right],\label{eq:H_K_transformed}\\
\tilde{\mathcal{H}}_{XXZ} & =\sum_{i}J_{H}^{z}S_{i}^{z}S_{i+1}^{z}+\nonumber \\
 & +\sum_{i}\frac{J_{H}^{\perp}}{2}e^{i\gamma\left[\phi_{i,s}^{R}\left(0\right)-\phi_{i+1,s}^{R}\left(0\right)\right]}S_{i}^{+}S_{i+1}^{-}+\text{H.c.},\label{eq:H_xxz_transformed}
\end{align}
 where we have defined $\tilde{\delta}_{s}\equiv\delta_{s}-\pi\gamma/2\sqrt{2}$.
Note that in the transformed representation, the quantum dynamics
of the bath {[}represented by the chiral field $\phi_{i,s}^{R}\left(0\right)${]}
appears explicitly in the Heisenberg term $\sim J_{H}^{\perp}\left(S_{i}^{+}S_{i+1}^{-}e^{i\gamma\left[\phi_{i,s}^{R}\left(0\right)-\phi_{i+1,s}^{R}\left(0\right)\right]}+\text{H.c.}\right)$.
Physically, this means that the Heisenberg interaction is now ``dressed''
by the spin-density fluctuations of the electron gas. %
{} Note that the independent-bath model (\ref{eq:H_F_independent_bath})
is crucial to implement bosonization along the chains, and to put
these ideas on a clear mathematical framework.

Up to now the parameter $\gamma$ appearing in Eq. (\ref{eq:transf_U})
remains completely arbitrary. We now set $\gamma=\sqrt{2}$ in Eqs.
(\ref{eq:H_K_transformed}) and (\ref{eq:H_xxz_transformed}), and
the transformed Hamiltonians simplify to 
\begin{align}
\tilde{\mathcal{H}}_{K} & =\sum_{i}-\frac{2\tilde{\delta}_{s}}{\pi\rho_{0}}S_{i}^{z}\frac{\nabla\phi_{i,s}^{R}\left(0\right)}{\sqrt{2}\pi}+\frac{J_{K}^{\perp}}{2\pi}S_{i}^{x},\label{eq:H_K_transformed_gamma_sqrt2}\\
\tilde{\mathcal{H}}_{XXZ} & =\sum_{i}J_{H}^{z}S_{i}^{z}S_{i+1}^{z}\nonumber \\
 & +\frac{J_{H}^{\perp}}{2}e^{i\sqrt{2}\left[\phi_{i,s}^{R}\left(0\right)-\phi_{i+1,s}^{R}\left(0\right)\right]}S_{i}^{+}S_{i+1}^{-}+\text{H.c.},\label{eq:H_xxz_transformed_gamma_sqrt2}
\end{align}
 where now $\tilde{\mathcal{H}}_{K}$ is equivalent to the spin-boson
model with Ohmic dissipation,\cite{Guinea85_Bosonization_of_a_two_level_system,leggett_two_state,Weiss99_Quantum_Dissipative_Systems}
with $\tilde{\delta}_{s}$ related to the dissipative parameter $\alpha$
in the context of macroscopic quantum coherence through $\alpha=\left(2\tilde{\delta}_{s}/\pi\right)^{2}$,
and with the in-plane Kondo interaction playing the role of a magnetic
field along the $x-$axis $h_{x}=-J_{K}^{\perp}/2\pi$%
. That model describes a quantum phase transition from a phase with
a ``frozen'' spin state (either $S_{i}^{z}=+1/2$ or $S_{i}^{z}=-1/2$)
for $\alpha>1$, to a phase with an ``untrapped'' spin state for
$\alpha<1$, where spin-flips induced by $S_{i}^{x}$ proliferate.
The strong-coupling regime of the single-impurity Kondo Hamiltonian
therefore corresponds to this last case, where (not too close to the
transition) the Kondo temperature is\cite{Guinea85_Bosonization_of_a_two_level_system,Guinea97_Quantum_dissipative_systems_review,LeHur08_Dissipative_Two-level_systems_review}
\begin{align}
T_{K} & \propto J_{K}^{\perp}\left(\frac{J_{K}^{\perp}b_{0}}{v_{F}}\right)^{\alpha/\left(1-\alpha\right)}.\label{eq:TK}
\end{align}

The special case $\alpha=0$ (i.e., $\tilde{\delta}_{s}=0$) was analyzed
by Kotliar and Si in Ref. {[}\onlinecite{Kotliar96_Toulouse_points_in_the_generalized_Anderson_model}{]}
and represents a particular limit where $\tilde{\mathcal{H}}_{K}$
can be diagonalized in the eigenbasis $\left|\pm\right\rangle _{i}$
of the operator $S_{i}^{x}$ (i.e., ``Toulouse point II'' in Ref.{[}\onlinecite{Kotliar96_Toulouse_points_in_the_generalized_Anderson_model}{]}).
Note that the condition $\tilde{\delta}_{s}=0$ implies that the original
phase-shift is $\delta_{s}=\pi/2$, corresponding to the unitary limit
$J_{K}^{z}b_{0}/v_{F}\rightarrow\infty$. Unfortunately, the unitary
limit is not consistent with the well-known local Fermi-liquid description
of the strong-coupling Kondo fixed point.\cite{nozieres74} Intuitively,
in the limit $\tilde{\delta}_{s}=0$ the coupling to the bath vanishes
and $\tilde{\mathcal{H}}_{K}$ reduces to a Zeeman Hamiltonian, which
is not equivalent to the Kondo problem.\cite{giamarchi_book_1d} However,
as shown by Kotliar and Si, a physically correct description of the
strong-coupling limit is recovered by performing second-order perturbation
expansion in $\tilde{\delta}_{s}$.\cite{Kotliar96_Toulouse_points_in_the_generalized_Anderson_model}
Physically, this is equivalent to reintroducing the coupling to the
bath.

In the case of plane-anisotropy, the Kondo couplings satisfy $J_{K}^{\perp}>J_{K}^{z}$.
This implies that in order to perform an expansion around the point
$\tilde{\delta}_{s}=0$, the microscopic parameters of the model should
be in a regime such that strictly speaking the use of bosonization
is not justified (i.e., the interactions are of the order or bigger
than the Fermi energy). However, since this approach has been shown
to successfully capture qualitatively the main features of the strong-coupling
Kondo fixed point,\cite{Kotliar96_Toulouse_points_in_the_generalized_Anderson_model}
we expect our approach to be correct only at a qualitative level.
In addition, since $\tilde{\delta}_{s}\approx0$ implies $\alpha\approx0$,
we will use Eq. (\ref{eq:TK}) to identify $T_{K}\simeq J_{K}^{\perp}$
in what follows.

An effective low-energy Hamiltonian valid near $\tilde{\delta}_{s}=0$,
and for the case where the Kondo interaction dominates over Heisenberg
exchange, i.e., $J_{K}^{\perp}\gg\left\{ J_{H}^{\perp},J_{H}^{z}\right\} $
can be obtained expanding Eq. (\ref{eq:H_K_transformed_gamma_sqrt2})
at order $\tilde{\delta}_{s}^{2}$ and projecting onto the lowest
energy level on each site $\left|-\right\rangle _{i}$, 
\begin{align}
\tilde{\mathcal{H}}_{K}^{\prime} & \equiv P_{-}\tilde{\mathcal{H}}_{K}P_{-}=-\sum_{i}\frac{1}{4}\left(\frac{2\tilde{\delta}_{s}}{\pi\rho_{0}}\right)^{2}\left(\frac{2\pi}{J_{K}^{\perp}}\right)\left(\frac{\nabla\phi_{i,s}^{R}\left(0\right)}{\sqrt{2}\pi}\right)^{2},\label{eq:H_K_projected}
\end{align}
 where we have introduced the projector operator on the subspace spanned
by $\left|-\right\rangle _{i}$ , i.e., $P_{-}\equiv\prod_{i}\left(\left|-\right\rangle _{i}\left\langle -\right|_{i}\right)$,
and where we have neglected a constant term $J_{K}^{\perp}/2\pi$.
The effective magnetic field $h_{x}$ opens a gap of size $\Delta=h_{x}$
in the spin excitation spectrum and consequently the spin degrees
of freedom are ``frozen'' in the lowest energy configuration $\left|-\right\rangle _{i}$.
In contrast, spin-density fluctuations in the bath remain gapless
and their dynamics becomes dominant at low energies. Projecting $\tilde{\mathcal{H}}_{XXZ}$
onto this basis yields 
\begin{align}
\tilde{\mathcal{H}}_{XXZ}^{\prime} & \equiv P_{-}\tilde{\mathcal{H}}_{XXZ}P_{-}\nonumber \\
 & =\sum_{i}\frac{J_{H}^{\perp}}{4}\cos\sqrt{2}\left[\phi_{i,s}^{R}\left(0\right)-\phi_{i+1,s}^{R}\left(0\right)\right].\label{eq:H_xxz_projected}
\end{align}
 In this representation, the Heisenberg term $J_{H}^{\perp}\left(S_{i}^{+}S_{i+1}^{-}+\text{h.c.}\right)$
induces an effective interaction between neighboring baths, encoded
in the term$\sim\cos\sqrt{2}\left[\phi_{i,s}^{R}\left(0\right)-\phi_{i+1,s}^{R}\left(0\right)\right]$.
This is an important result in our work, complementary to the situation
analyzed in Sec. \ref{sub:weak_coupling_RG}, where the opposite limit
$J_{H}^{\perp}\gg J_{K}^{\perp}$ was studied. In that case, the bath
was integrated out, and we studied the stability of the TLL phase
to leading order in perturbation theory. Here, we do just the opposite:
we keep the degrees of freedom of the bath and eliminate the spin
degrees of freedom.

Although a general derivation of an effective low-energy model (i.e.,
for an arbitrary value of $\gamma$), is beyond the scope of the present
work, it is worth noting that the effective coupling between fermionic
baths that appears in Eq. (\ref{eq:H_xxz_projected}) is a general
physical feature that does not depend on our particular derivation
for $\gamma=\sqrt{2}$. This can be understood using, for example,
similar arguments as those leading to the Nozi{\`e}res' local Fermi-liquid.\cite{nozieres74}
Indeed, when $J_{K}^{\perp}\gg J_{H}^{\perp}$, a natural approach
is to start from the Kondo singlets at neighboring sites $i$ and
$i+1$, i.e., $\left|G_{i}\right\rangle $ and $\left|G_{i+1}\right\rangle $
respectively, where$\left|G_{l}\right\rangle =\left(\left|\Uparrow\right\rangle _{l}\left|c_{l0,\downarrow}^{\dagger}\right\rangle -\left|\Downarrow\right\rangle _{l}\left|c_{l0,\uparrow}^{\dagger}\right\rangle \right)/\sqrt{2}$.
The perturbation $\mathcal{H}^{\prime}=-t\sum_{l=\left\{ i,i+1\right\} }\left(c_{l1,\sigma}^{\dagger}c_{l0,\sigma}+\text{h.c.}\right)$
acting on these neighboring singlets produces virtual excitations
to the $n_{l0}=1$ triplet subspace, at order $\left(t/J_{K}^{\perp}\right)^{2}$
on each one of them. Eventually, the Heisenberg interaction $J_{H}^{\perp}\left(S_{i}^{+}S_{i+1}^{-}+\text{h.c.}\right)$
restores the initial singlet ground states $\left|G_{i}\right\rangle $
and $\left|G_{i+1}\right\rangle $ and, as a net result, virtual processes
generate an effective spin interaction $\sim J_{H}^{\perp}\left(t/J_{K}^{\perp}\right)^{4}\left(c_{i,1,\uparrow}^{\dagger}c_{i+1,1,\downarrow}+\text{h.c.}\right)$
between the second sites in the chains $i$ and $i+1$. Bosonizing
this induced effective interaction yields a term of the form $\sim\cos\sqrt{2}\left[\phi_{i,s}^{R}\left(0\right)-\phi_{i+1,s}^{R}\left(0\right)\right]$,
analogous to Eq. (\ref{eq:H_xxz_projected}).

In order to derive an effective 1D model, we integrate out of the
modes $\phi_{i,s}^{R}\left(y\right)$ for $y\neq0$. This can be done
exactly using the functional integral representation of the partition
function, and generates a term of the form $\sim\left|\omega_{m}\right|$
in the effective action, which stems from the (Ohmic) dissipation
induced by the coupling to local bath (cf. Refs.~{[}\onlinecite{giamarchi_book_1d,gogolin_1dbook}{]},
for details). The resulting Euclidean action of the system reads 
\begin{align}
\mathcal{S}^{\prime} & =\mathcal{S}_{0}^{\prime}+\mathcal{S}_{H}^{\prime},\label{eq:action_strong_coupling}\\
\mathcal{S}_{0}^{\prime} & =\sum_{i}\left[\sum_{\omega_{m}}\frac{\left|\omega_{m}\right|}{4\pi\beta}\left|\varphi_{i}\left(i\omega_{m}\right)\right|^{2}+\int_{0}^{\beta}d\tau\frac{\left(\partial_{\tau}\varphi_{i}\left(\tau\right)\right)^{2}}{2\pi E_{0}}\right]\label{eq:S_0}\\
\mathcal{S}_{H}^{\prime} & =\sum_{i}\int_{0}^{\beta}d\tau\;\frac{J_{H}^{\perp}}{4}\cos\left[\varphi_{i}\left(\tau\right)-\varphi_{i+1}\left(\tau\right)\right],\label{eq:S_H}
\end{align}
 where we have defined the more compact notation for local field $\varphi_{i}\equiv\sqrt{2}\phi_{i,s}^{R}\left(y=0\right)$,
and where we have used the equation of motion of chiral fields, i.e.,
$\partial_{\tau}\varphi_{i}-iv_{F}\nabla\varphi_{i}=0$ to express
$\nabla\varphi_{i}$ in terms of $\partial_{\tau}\varphi_{i}$. In
addition, in Eq. (\ref{eq:S_0}) we have defined the parameter%

\begin{align}
E_{0} & \equiv\frac{J_{K}^{\perp}}{4\tilde{\delta}_{s}^{2}},\label{eq:E_0}
\end{align}
 where the singularity at $\tilde{\delta}_{s}=0$ is a consequence
of the unphysical unitary limit mentioned above.

Note that the effective model Eq. (\ref{eq:action_strong_coupling})
is formally equivalent to the action of a 1D Josephson-junction array
with local Ohmic dissipation, with $\varphi_{i}$ the phase of the
superconducting order parameter at site $i$, $J_{H}^{\perp}$ the
Josephson coupling and $E_{0}$ the charging energy with respect to
the ground.\cite{Lobos12_comment_1D_Kondo_cloud,Tewari06_Dissipate_locally_couple_globally}
It can be also brought to a form equivalent to a 1D O(2) dissipative
quantum rotor model if we write it in terms of $\mathbf{N}_{i}\left(\tau\right)=\left(\cos\varphi_{i}\left(\tau\right),\sin\varphi_{i}\left(\tau\right)\right)$.\cite{cazalilla06_dissipative_transition,drewes03_dissipative_jj,Renn97_condmat_dissipative_quantum_rotors,Werner05_Quantum_Spin_Chains_with_site_dissipation}
The fact that Hamiltonian Eq. (\ref{eq:H_total}) can be mapped (in
the limit $J_{K}^{\perp},J_{K}^{z}\gg J_{H}^{\perp},J_{H}^{z}$) to
these dissipative models is an important result of our work which
shows interesting underlying connections between apparently different
physical situations. %
{} %

To appreciate the physical consequences of the effective model in
Eq.~\eqref{eq:action_strong_coupling}, we concentrate on the transverse
spin correlation function $\mathcal{C}^{+-}\left(n,\tau\right)=\left\langle T_{\tau}S_{i+n}^{+}\left(\tau\right)S_{i}^{-}\left(0\right)\right\rangle $.
In the transformed representation, this correlation is evaluated as
(cf. Appendix \ref{sec:spin_correlation_function}) $\mathcal{C}^{+-}\left(n,\tau\right)=\left\langle T_{\tau}e^{i\varphi_{i+n}\left(\tau\right)}e^{-i\varphi_{i}\left(0\right)}\right\rangle $.
Near the strong-coupling single-impurity Kondo limit $J_{K}^{\perp}\gg\left\{ J_{H}^{\perp},J_{H}^{z}\right\} $,
we obtain the result (cf. Eq. \ref{eq:C_Kondo_limit}) $\mathcal{C}^{+-}\left(0,\tau\right)\sim\tau^{-2}$,
as expected for the local Fermi-liquid description of the Kondo problem.\cite{nozieres74,Kotliar96_Toulouse_points_in_the_generalized_Anderson_model}
This slow decay is a consequence of the Ohmic dissipation term $\sim\left|\omega_{m}\right|$
in Eq. (\ref{eq:S_0}), inherited from the dynamics of the semi-infinite
fermionic chain. This behavior is consistent with the strong-coupling
Kondo picture where, at long times, the spin degrees of freedom are
merged with those of the fermion bath.\cite{Kotliar96_Toulouse_points_in_the_generalized_Anderson_model}
On the other hand, the exponentially decaying non-local correlation
{[}cf. Eq. (\ref{eq:C_non_local_static}){]} 
\begin{align}
\mathcal{C}_{n}^{+-}\left(n,0\right) & =\frac{1}{2}\frac{e^{-n/\xi_{c}}}{n+1},\label{eq:C_non_local_Kondo}
\end{align}
 with $\xi_{c}\equiv1/\ln\left|\frac{8e^{\gamma_{E}}}{\pi}\frac{E_{0}}{J_{H}^{\perp}}\right|$the
correlation length (where $\gamma_{E}=0.577\dots$ is the Euler gamma
constant), indicates that the spins are not spatially correlated beyond
a distance $\xi_{c}$, supporting the idea that in this limit the
spin chain realizes a disordered phase of nearly independent Kondo
singlets.%

\subsubsection{Hubbard-Stratonovich decoupling and RG analysis of the effective
$\psi^{4}-$theory}

\begin{figure}
\includegraphics[scale=0.8]{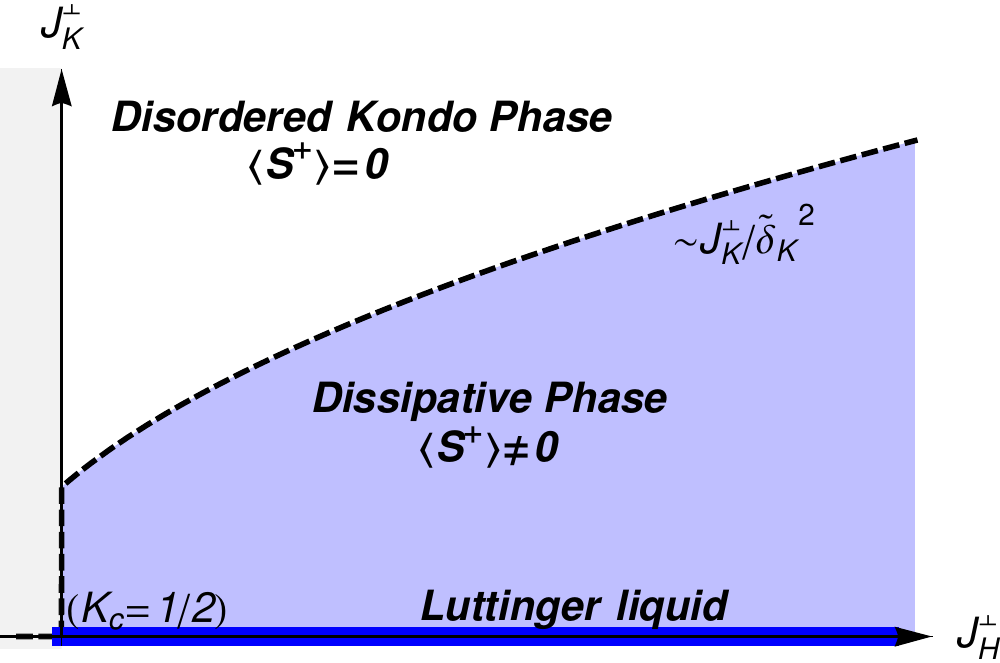}

\caption{Schematic $T=0$ phase diagram of the dissipative 1D Kondo-Heisenberg
spin chain. The thick blue line at $J_{K}^{\perp}=0$ corresponds
to the Tomonaga-Luttinger liquid phase, which is unstable against
a small perturbation $J_{K}^{\perp}$. The dashed line is the critical
line separating a phase of disordered Kondo-singlets (upper white
area), from a dissipative phase (shaded bottom area), characterized
by gapless diffusive spin excitations. In this phase, at $T=0$, the
dissipative dynamics induced by the metallic environment stabilizes
long-range order in the transverse magnetization. \label{fig:phase_diagram}}
\end{figure}

The properties and phases of action (\ref{eq:action_strong_coupling})
can be investigated introducing an auxiliary bosonic field $\psi_{i}\left(\tau\right)$
to decouple the Heisenberg term $J_{H}^{\perp}$ (cf. e.g. Refs. {[}\onlinecite{feigelman98_smt_in_2d_proximity_array,sachdev_book,Lutchyn08_Dissipative_QPT_in_SC-graphene_systems}{]})\begin{widetext}
\begin{align}
\frac{J_{H}^{\perp}}{4}\sum_{i}\int_{0}^{\beta}d\tau\;\cos\left[\varphi_{i}\left(\tau\right)-\varphi_{i+1}\left(\tau\right)\right]\rightarrow & \int_{0}^{\beta}d\tau\sum_{i,j}\psi_{i}^{*}\left(\tau\right)\left[\mathbf{J^{-1}}\right]_{ij}\psi_{j}\left(\tau\right)-\frac{1}{2}\int_{0}^{\beta}d\tau\sum_{i}\left[\psi_{i}^{*}\left(\tau\right)e^{i\varphi_{i}\left(\tau\right)}+\psi_{i}\left(\tau\right)e^{-i\varphi_{i}\left(\tau\right)}\right],
\end{align}
 \end{widetext} where we have defined the matrix $\left[\mathbf{J}\right]_{ij}\equiv\frac{1}{8}J_{H}^{\perp}\left(\delta_{i,j+1}+\delta_{i+1,j}\right)$.
Then, the partition function reads $Z=Z_{0}\int\mathcal{D}\left[\psi\right]\; e^{-S\left[\psi\right]}$,
where

\begin{align}
\mathcal{S}\left[\psi\right] & =\sum_{i,j}\int_{0}^{1/T}d\tau\;\psi_{i}^{*}\left(\tau\right)\left[\mathbf{J^{-1}}\right]_{ij}\psi_{j}\left(\tau\right)\nonumber \\
 & -\ln\left\langle \exp\frac{1}{2}\sum_{i}\int_{0}^{1/T}d\tau\;\left[\psi_{i}^{*}\left(\tau\right)e^{i\varphi_{i}\left(\tau\right)}+\text{h.c.}\right]\right\rangle _{0},\label{eq:S_psi}
\end{align}
 is the effective action for the auxiliary field $\psi_{i}\left(\tau\right)$.
Here, the notation $\left\langle \dots\right\rangle _{0}$ means average
with respect to the local action (\ref{eq:S_0}). A cumulant expansion
of the last term in Eq. (\ref{eq:S_psi}) to order $\psi^{4}$ yields
\begin{align}
\mathcal{S}\left[\psi\right] & =\frac{T}{2N_{i}}\sum_{\mathbf{q}}G_{0}^{-1}\left(\mathbf{q}\right)\left|\psi_{\mathbf{q}}\right|^{2}+\frac{u}{4!}\sum_{i}\int_{0}^{1/T}d\tau\;\left|\psi_{i}\left(\tau\right)\right|^{4},\label{eq:S_4th_order}
\end{align}
 where the compact notation $\mathbf{q}=\left(k,\omega_{m}\right)$,
with $\omega_{m}\equiv2\pi mT$ the bosonic Matsubara frequencies,\cite{mahan2000}
has been used, and where $N_{i}$ is the number of spins. Here we
have defined the Gaussian propagator

\begin{align}
G_{0}^{-1}\left(\mathbf{q}\right) & \equiv r+\frac{4}{J_{H}^{\perp}}k^{2}+\frac{\pi e^{-2\gamma_{E}}}{E_{0}^{2}}\left|\omega_{m}\right|,\label{eq:G_0}
\end{align}
 where 
\begin{align}
r & \equiv\frac{8}{J_{H}^{\perp}}-\frac{1}{2}\int_{0}^{\beta}d\tau\; e^{i\omega_{m}\tau}\left\langle T_{\tau}e^{i\Phi_{m}\left(\tau_{1}\right)}e^{-i\Phi_{m}\left(\tau_{2}\right)}\right\rangle _{0}\label{eq:r_parameter}\\
 & =\frac{8}{J_{H}^{\perp}}-\frac{\pi e^{-\gamma_{E}}}{2E_{0}}\\
u & \equiv\frac{c}{E_{0}^{3}},\label{eq:u_parameter}
\end{align}
 with $c=21.8\dots$ a numerical coefficient. At the mean-field level
{[}i.e., the saddle-point approximation to Eq.~(\ref{eq:S_4th_order}){]},
this model describes a QPT when $J_{H}^{\perp}$ reaches the critical
value $J_{H,c}^{\perp}\equiv16E_{0}/\pi e^{-\gamma_{E}}\propto J_{K}^{\perp}/\tilde{\delta}_{s}^{2}$.
In Fig. \ref{fig:phase_diagram} we show schematically this critical
line as a dashed black line, separating the disordered phase with
$\left\langle \psi_{i}\right\rangle =0$, corresponding to the disordered
Kondo phase described in Section \ref{sub:strong_coupling} {[}cf.
Eq. (\ref{eq:C_non_local_Kondo}){]}, from the ordered phase with
$\left\langle \psi_{i}\right\rangle \neq0$. Physically, when $J_{H}^{\perp}>J_{H,c}^{\perp}$
the Heisenberg interaction is large enough to induce long-range coherence
in the transverse magnetization $\left\langle S_{i}^{+}\right\rangle \propto\left\langle e^{i\varphi_{i}}\right\rangle \propto\left\langle \psi_{i}\right\rangle $
along the spin-chain. On the other hand, when $J_{H}^{\perp}<J_{H,c}^{\perp}$
the ``charging'' term $E_{0}$ induces large quantum fluctuations
of the field $\varphi_{i}$ (i.e., local spin-flips induced by the
term $J_{K}^{\perp}$) and therefore tends to destroy the ordered
state. In this representation, the competition between the couplings
$J_{K}^{\perp}$ and $J_{H}^{\perp}$ becomes transparent.

Beyond the mean-field level, the quantum critical properties of this
model, generalized to describe a $N-$component field $\left\{ \psi_{\alpha}\left(x\right)\right\} =\left\{ \psi_{1}\left(x\right),\psi_{2}\left(x\right),\dots,\psi_{N}\left(x\right)\right\} $
in $d$ dimensions, are well-known and have been studied in the context
of antiferromagnetic instabilities of Fermi-liquids,\cite{sachdev_book,Pankov04_NFL_behavior_and_2D_AFM_fluctuations}
using the framework of the Hertz-Moriya-Millis theory.\cite{Hertz76_Quantum_Critical_Phenomena,Moriya73_Critical_phenomena,Millis93_Quantum_critical_points_in_Fermi_systems}
This theory describe quantum fluctuations of the order parameter $\psi_{\alpha}\left(x\right)$,
with critical dynamical exponent $z=2$, around the Gaussian fixed-point
{[}i.e., $u=0$ in Eq. (\ref{eq:S_4th_order}){]}. A standard momentum-shell
RG procedure, performing a two-loop expansion in $u$, and in the
small parameter $\epsilon=4-D$, where $D=d+z$, allows to obtain
the RG-flow equations for the parameters of model (\ref{eq:S_4th_order}),
\cite{sachdev_book,Pankov04_NFL_behavior_and_2D_AFM_fluctuations}
{} 
\begin{align}
\frac{dr\left(\ell\right)}{d\ell} & =2r\left(\ell\right)+u\left(\ell\right)\frac{N+2}{6}\frac{A_{D}}{1+r\left(\ell\right)},\label{eq:RG_r}\\
\frac{du\left(\ell\right)}{d\ell} & =\epsilon u\left(\ell\right)-u^{2}\left(\ell\right)\frac{N+8}{6}\frac{A_{D}}{\left(1+r\left(\ell\right)\right)^{2}},\label{eq:RG_u}
\end{align}
 where $A_{D}=2\pi^{\left(D/2\right)}/\Gamma\left(d/2\right)$ (with
$\Gamma\left(x\right)$ the Euler gamma function) is the surface area
of the $D-$dimensional sphere, and where the units are such that
the high-energy cutoff of the theory (\ref{eq:S_4th_order}) is $\Lambda=1$.
For $\epsilon$ small and positive ($d<2$), this RG-flow is controlled
by a Wilson-Fisher fixed-point located at $r^{*}=-\epsilon\left(N+2\right)/2\left(N+8\right)$
, $u^{*}=6\epsilon/A_{D}\left(N+8\right)$. %

The stability of the Wilson-Fisher fixed point can be studied upon
expansion of Eqs. (\ref{eq:RG_r}) and (\ref{eq:RG_u}) in the small
deviations $\delta r=r-r^{*}$, $\delta u=u-u^{*}$, and allows to
obtain the eigen-coupling equation $dw_{i}/d\ell=\lambda_{i}w_{i}$,
$\left(i=1,2\right)$, with eigenvalues $\lambda_{1}=2-\epsilon\left(N+2\right)/\left(N+8\right)+\mathcal{O}\left(\epsilon^{2}\right)$
and $\lambda_{2}=-\epsilon+\mathcal{O}\left(\epsilon^{2}\right)$,
determining the critical exponents of the transition.\cite{sachdev_book,Pankov04_NFL_behavior_and_2D_AFM_fluctuations}
Although for $d=1$ the parameter $\epsilon=4-(d+z)=1$, which is
not small, recent quantum Monte Carlo simulations \cite{Werner05_Quantum_Spin_Chains_with_site_dissipation}
have shown evidence of a QPT in XY-spin chains subject to local Ohmic
dissipation, with critical exponents in good agreement with those
predicted using the the $\epsilon$-expansion.~\cite{Pankov04_NFL_behavior_and_2D_AFM_fluctuations,Werner05_Quantum_Spin_Chains_with_site_dissipation}

In the ordered phase $\left\langle \psi_{i}\right\rangle \neq0$ occurring
for $J_{H}^{\perp}>J_{H,c}^{\perp}$, since the U(1) symmetry of Eq.
(\ref{eq:S_psi}) is spontaneously broken, a Goldstone mode arises
from smooth fluctuations of the phase of the order parameter $\psi_{i}\simeq\left|\psi_{i}\right|e^{i\vartheta_{i}}$,
and it becomes necessary to check the stability of the ordered-phase.
To that end, we return to Eq. (\ref{eq:S_4th_order}) and perform
an expansion in small fluctuations of the phase $\delta\vartheta_{i}=\vartheta_{i}-\vartheta_{0}$,
around an arbitrary value $\vartheta_{0}$. At Gaussian order in $\delta\vartheta_{i}$
we obtain the effective action 
\begin{align}
\mathcal{S}_{\text{eff}}\left[\vartheta\right] & =\frac{1}{2}\frac{1}{\beta N_{i}}\sum_{\mathbf{q}}G_{\text{eff}}^{-1}\left(\mathbf{q}\right)\left|\vartheta\left(\mathbf{q}\right)\right|^{2},\label{eq:S_Goldstone_mode}\\
G_{\text{eff}}^{-1}\left(\mathbf{q}\right) & =\frac{\pi e^{-2\gamma_{E}}\psi_{0}^{2}}{E_{0}^{2}}\left|\omega_{m}\right|+\frac{4\psi_{0}^{2}}{J_{H}^{\perp}}k^{2},\label{eq:propagator_Goldstone_mode}
\end{align}
 where $G_{\text{eff}}^{-1}\left(\mathbf{q}\right)$ is the propagator
of the Goldstone mode, and where $\psi_{0}$ is the saddle-point solution
of Eq. (\ref{eq:S_4th_order}). This propagator describes a gapless
phase characterized by diffusive ($z=2$) excitations of the field
$\vartheta\left(\mathbf{q}\right)$, and by correlation functions
$\mathcal{C}^{+-}\left(x,\tau\right)\equiv\left\langle T_{\tau}S^{+}\left(x,\tau\right)S^{-}\left(0\right)\right\rangle $
decaying as $\mathcal{C}^{+-}\left(x,0\right)\sim\left|x\right|^{-1}$
at long distances, and $\mathcal{C}^{+-}\left(\tau\right)\sim\left|\tau\right|^{-1/2}$
at long times (cf. Appendix \ref{sub:spin_correlator_dissipative}).
Using Eq. (\ref{eq:S_Goldstone_mode}) to evaluate the average of
the order parameter $\left\langle \psi_{i}\right\rangle =\psi_{0}\left\langle e^{i\vartheta_{i}}\right\rangle =\psi_{0}e^{-\frac{1}{2}\left\langle \vartheta_{i}^{2}\right\rangle }$,
we obtain the result 
\begin{align}
\left\langle \psi_{i}\right\rangle  & =\psi_{0}\exp\left[-\left(\frac{e^{\gamma_{E}}}{2\pi}\right)^{2}\frac{2E_{0}}{\psi_{0}^{2}}\sqrt{J_{H}^{\perp}E_{0}}\right] & \quad\left(\text{at }T=0\right).
\end{align}
 Interestingly, due to the presence of the dissipative term $\sim\left|\omega_{m}\right|$,
the (Gaussian) fluctuations of the spin-chain are strongly suppressed
relative to the isolated (XY) chain. Indeed, contrary to the case
of isolated 1D systems, quantum fluctuations do not destroy the LRO
because the effective dimensionality of the quantum system is $D=d+z=3$,
larger than the critical dimension $D_{c}=2$ determined by the Gaussian
theory for the Goldstone mode Eq. (\ref{eq:S_Goldstone_mode}). Note
that this is not in contradiction with the Mermin-Wagner theorem,\cite{mermin_wagner_theorem}
which predicts the destruction of LRO at $T=0$ in 1D systems with
short range interactions. In our case, due to the presence of a higher
dimensional fermionic bath which induces long-ranged (imaginary) time
correlations, the system cannot be considered strictly one dimensional.
Therefore, the Mermin-Wagner theorem is not applicable to our system.

At this point, it is interesting to note the connection with the weak-coupling
TLL description of Sec. \ref{sub:weak_coupling_RG}, which becomes
apparent using, for instance, the self-consistent harmonic approximation
(SCHA) method.\cite{feynman_statmech} This method consists in finding
the optimal propagator $G_{\text{trial}}^{-1}\left(\mathbf{q}\right)$
of a trial Gaussian action, such that the variational free-energy
of the system is minimized. For the case of Sec. \ref{sub:weak_coupling_RG},
when $K>1/2$ the only relevant variable is the field $\Theta\left(x\right)$,
describing the transverse spin-excitations of the spin-chain weakly
coupled to the metal, and consequently the trial action writes compactly
as $S_{\text{trial}}=\frac{1}{2\beta L}\sum_{\mathbf{q}}G_{\text{trial}}^{-1}\left(\mathbf{q}\right)\left|\Theta\left(\mathbf{q}\right)\right|^{2}$.
Here we do not show the derivation of the SCHA equations, and refer
the reader to Refs. {[}\onlinecite{cazalilla06_dissipative_transition, lobos09_dissipation_scwires}{]},
where this method was applied to closely related 1D TLL systems in
contact to dissipative baths. The optimal propagator is $G_{\text{trial}}^{-1}\left(\mathbf{q}\right)=\eta\left|\omega_{m}\right|+K\left(uk^{2}+\omega_{m}^{2}/u\right)/\pi$,
with $\eta\sim J_{K}^{\perp}$. The similar forms (in the limit $\mathbf{q}\rightarrow0$)
of $G_{\text{trial}}^{-1}\left(\mathbf{q}\right)$ and $G_{\text{eff}}^{-1}\left(\mathbf{q}\right)$
in Eq. (\ref{eq:propagator_Goldstone_mode}) suggests that the dissipative
gapless phase with $z=2$, obtained in the strong-coupling regime,
is also stable in the weak-coupling regime (see lower part of Fig.
\ref{fig:phase_diagram}). The thick solid line at the bottom (i.e.,
$J_{K}^{\perp}=0$) corresponds to the gapless TLL phase described
by the Hamiltonian (\ref{eq:H_XXZ_Bosonized}). As shown by the RG-flow
Eq. (\ref{eq:RG_g_perps}), this line is unstable against a vanishingly
small perturbation $J_{K}^{\perp}$.

\section{\label{sec:summary}Summary and conclusions}

In this work we have studied the quantum critical properties of a
$S=1/2$ spin chain described by the anisotropic XXZ Hamiltonian (\ref{eq:H_xxz}),
coupled to a metallic environment via the anisotropic Kondo model
(\ref{eq:H_K}). From the theoretical point of view, this model is
related to the well-known Kondo-lattice model, which is relevant to
the description of (higher dimensional) heavy-fermion systems.\cite{hewson}
In addition, it also encodes an interesting interplay between quantum
spin-fluctuations, enhanced by the one-dimensional geometry, and dissipation
due to the Kondo coupling to the metallic environment. From the point
of view of experimental systems, this is a relevant model for the
description of linear arrangements of magnetic impurities built on
the top of clean metallic surfaces by the means of STM techniques
(see Fig. \ref{fig:system-1}).\cite{Hirjibehedin06_Mn_spin_chains,Loth12_Bistability_in_atomic_scale_AFM}

We study the model Eq. (\ref{eq:H_total}) in different regimes of
parameters using various analytical approaches (i.e., Abelian bosonization,
renormalization-group method, analysis of the Ginzburg-Landau functional,
etc.), and obtain the quantum phases at $T=0$.

There are two crucial assumptions in our work: A)The presence of easy-plane
anisotropy in $\mathcal{H}_{XXZ}$ and $\mathcal{H}_{K}$, that favors
the couplings in the plane of the metallic surface. In real systems,
this assumption is physically reasonable due to the presence of Dzyaloshinskii-Moriya
interactions (induced by spin-orbit coupling at the surface of metals)
which break the SU(2) invariance, and enhance the transverse Heisenberg
coupling $J_{H}^{\perp}$ {[}cf. Eq. (\ref{eq:transf_Spm}){]}.\cite{Carbone11_Self-assembled_magnetic_networks_on_surfaces,Garate10_Anisotropic_spin_chains}
Other physical mechanisms producing other types of anisotropy, such
as magnetocrystalline effects in the metallic host are beyond the
scope of the present work.\cite{Hirjibehedin07_Magnetic_anisotropy_in_surfaces,Barral10_Anisotropic_Kondo}
B) The different dimensionality between the spin chain and the metallic
environment, which allows to assume the presence of enough Kondo-screening
conduction electrons per spin in the chain. 

In Sec. \ref{sub:weak_coupling_RG} we set the stage by investigating
the weak-coupling regime $J_{K}^{\perp}\ll J_{H}^{\perp}$ (see Sec.
\ref{sub:weak_coupling_RG}), where $\mathcal{H}_{K}$ can be considered
a perturbation to the isolated spin-chain. In that case, we derive
a set of RG-flow equations {[}cf. Eqs. (\ref{eq:RG_g_zu})-(\ref{eq:RG_g_perps}){]}
at first order in the parameters $J_{K}^{\perp}$ and $J_{K}^{z}$.
This RG-flow is determined by the value of the Luttinger parameter
$K$ of the chain, and is drastically different the RG-flow expected
for the single-impurity Kondo problem.\cite{hewson} Far away from
the isotropic SU(2) point $K=1/2$, while $J_{K}^{z}$ becomes an
irrelevant coupling (in the RG sense), $J_{K}^{\perp}$ becomes relevant
and destabilizes the Tomonaga-Luttinger liquid fixed-point (cf. thick
bottom line in Fig. \ref{fig:phase_diagram}). This situation is analogous
to other TLL systems coupled to a dissipative environment.\cite{cazalilla06_dissipative_transition,lobos09_dissipation_scwires}
Closer to the symmetric point $K=1/2$, the effect of the irrelevant
coupling $J_{K}^{z}$ becomes more important and higher-order perturbation
analysis is required to study the RG-flow of parameters. An important
point about the scaling analysis of Sec. \ref{sub:weak_coupling_RG}
is that it indicates that that the most relevant contribution (in
the RG sense) to the free-energy $\Delta F$ comes from the local
sector of the spin-response of the metal (this is discussed in detail
in Appendix \ref{sec:RG_free-energy}). At this stage the only assumption
made about the bath is that is has not been affected by (weak) coupling
$J_{K}^{\perp}$.

In the opposite limit $J_{K}^{\perp}\gg J_{H}^{\perp}$ (cf. Sec.
\ref{sub:strong_coupling}), we consider the problem starting from
the limit of decoupled Kondo-screened impurities. Due to the higher
dimensionality of the metallic host as compared to the spin chain,
there are always sufficient conduction electrons to screen the impurity
spins and the exhaustion problem\cite{nozieres_exhaustion} never
appears in the present situation. This allowed us to perform a crucial
approximation, the ``independent bath approximation'', which amounts
to neglect interference between neighboring Kondo clouds and to consider
each spin as independently screened. Using this approximation, we
show that the effective model for coupled Kondo impurities {[}cf.
Eq. (\ref{eq:action_strong_coupling}){]} is formally equivalent to
that of a 1D Josephson-junction array with on-site Ohmic dissipation,
which is known to undergo a quantum phase transition as a function
of the dissipation parameter.\cite{Tewari06_Dissipate_locally_couple_globally}
To extract the properties of the spin chain in this limit, we derive
an effective Ginzburg-Landau theory {[}cf. Eq. (\ref{eq:S_4th_order}){]}
in terms of a U(1) bosonic Hubbard-Stratonovich field $\psi_{i}$,
and study the critical properties around the Gaussian fixed-point.\cite{sachdev_book,Pankov04_NFL_behavior_and_2D_AFM_fluctuations}
Physically, the order parameter $\psi_{i}$ describes LRO in the transverse
magnetization $\left\langle S_{i}^{+}\right\rangle \propto\left\langle \psi_{i}\right\rangle $
along the direction of the spin chain. In the limit $J_{K}^{\perp}\gg J_{H}^{\perp}$,
the system is characterized by a vanishing order parameter $\left\langle \psi_{i}\right\rangle =0$,
which we interpret as a manifestation of a disordered Kondo-singlet
phase, where the spins are screened by their local fermionic bath.
Upon the increase of $J_{H}^{\perp}$, the system experiences a QPT
(dashed line in Fig. \ref{fig:phase_diagram}) towards a phase with
LRO, characterized by a non-vanishing order-parameter $\left\langle \psi_{i}\right\rangle \neq0$
at $T=0$. Interestingly, the dynamics of the emerging Goldstone mode
{[}cf. Eq. (\ref{eq:S_Goldstone_mode}){]} is not able to destroy
the mean-field solution, which would be the usual situation for isolated
1D systems. This anomaly occurs due to the dissipative character of
this mode.

It would be very interesting to verify our predictions on experimental
level. Experimentally, the phase diagram could be studied with STM
techniques either by varying the strength of the Kondo exchange coupling
(i.e., by growing the 1D spin chain on the top of a decoupling layer
\cite{Hirjibehedin06_Mn_spin_chains}), or by changing the distance
between spins,\cite{Wahl07_Exchange_Interaction_between_Single_Magnetic_Adatoms}
which has the effect of changing the magnitude and sign of the exchange
interaction. More accurate predictions for the possible experimental
observables are at present under progress.\cite{Lobos12_unpublished_Kondo_Heisenberg}

\begin{acknowledgements}

The authors are grateful to T. Giamarchi, A. Georges, F. Guinea, M.
Fabrizio, A. Iucci and P. Simon for useful discussions. AML and PC
acknowledge support from the Swiss National Science Foundation under
MaNEP and Division II. AML acknowledges support from DARPA QuEST,
JQI-NSFPFC. MAC acknowledges the hospitality of D.-W. Wang at NCTS
(Taiwan) and A. H. Castro-Neto at the Graphene Research Center of
the National University of Singapore, and the support of Spanish MEC
through grant FIS2010-19609-C02-02.

\end{acknowledgements}

\appendix

\section{\label{sec:RG_free-energy}Leading corrections to the spin-chain
free energy}

We begin by considering the partition function 
\begin{align}
Z & =\text{Tr}\: e^{-\left(\mathcal{H}_{0}+\mathcal{H}_{K}\right)/T}=Z_{0}\left\langle T_{\tau}e^{-\int_{0}^{1/T}d\tau\mathcal{H}_{K}\left(\tau\right)}\right\rangle _{0},
\end{align}
 where $Z_{0}\equiv\text{Tr}\: e^{-\mathcal{H}_{0}/T}$ is the partition
function of the uncoupled host-chain system, and $\mathcal{H}_{0}=\mathcal{H}_{XXZ}+\mathcal{H}_{F}$
(cf. Eqs.~\ref{eq:H_XXZ_Bosonized} and \ref{eq:H_F}), and $\tau$
is the Matsubara imaginary time.~\cite{mahan2000} The average $\left\langle \mathcal{A}\right\rangle _{0}\equiv\text{Tr}\:\left[e^{-\mathcal{H}_{0}/T}\mathcal{A}\right]/\text{Tr}\:\left[e^{-\mathcal{H}_{0}/T}\right]$
stands for the thermodynamic average of an operator $\mathcal{A}$
over the Gibbs ensemble defined by $\mathcal{H}_{0}$.

The free-energy of the system (removing the bath contribution) is
given by $F-F_{0}=-T\ln(Z/Z_{0})=-T\ln\left\langle T_{\tau}e^{-\int_{0}^{1/T}d\tau\mathcal{H}_{K}\left(\tau\right)}\right\rangle _{0}$.
Upon expanding the exponential to the lowest non-trivial order, we
obtain the leading free-energy correction \begin{widetext} 
\begin{align}
\Delta F & =-T\ln\left[1+\frac{1}{2!}\int_{0}^{1/T}d\tau_{1}d\tau_{2}\left\langle T_{\tau}\mathcal{H}_{K}\left(\tau_{1}\right)\mathcal{H}_{K}\left(\tau_{2}\right)\right\rangle _{0}+\dots\right]=-\frac{T}{2}\int d\mathbf{r}_{1}d\mathbf{r}_{2}\left[\left(\frac{J_{K}^{z}}{k_{F}^{3}}\right)^{2}\left\langle T_{\tau}S^{z}\left(\mathbf{r}_{1}\right)S^{z}\left(\mathbf{r}_{2}\right)\right\rangle _{0}\chi^{zz}\left(\mathbf{r}_{12}\right)\right.\nonumber \\
 & \qquad\left.+2\left(\frac{J_{K}^{\perp}}{2k_{F}^{3}}\right)^{2}\left\langle T_{\tau}S^{+}\left(\mathbf{r}_{1}\right)S^{-}\left(\mathbf{r}_{2}\right)\right\rangle _{0}\chi^{-+}\left(\mathbf{r}_{12}\right)\right]+\cdots,\label{eq:Delta_F}
\end{align}
 \end{widetext} where we have introduced the compact notation $\mathbf{r}\equiv\left(x,\tau\right)$,
and $\mathbf{r}_{12}\equiv\mathbf{r}_{1}-\mathbf{r}_{2}$. We have
also defined the host spin-response function, $\chi^{ab}\left(\mathbf{R}_{1}-\mathbf{R}_{2},\tau_{1}-\tau_{2}\right)\equiv\left\langle T_{\tau}s^{a}\left(\mathbf{R}_{1},\tau_{1}\right)s^{b}\left(\mathbf{R}_{2},\tau_{2}\right)\right\rangle _{0}-\chi_{0}^{ab}\left(\mathbf{R}_{1}-\mathbf{R}_{2}\right)$,
$\chi_{0}^{ab}\left(\mathbf{R}_{1}-\mathbf{R}_{2}\right)$ being the
static part of the spin-response of the host, which has been already
included in the static spin-chain exchange couplings $J_{H}^{\perp}$
and $J_{H}^{z}$, determining the Luttinger parameter $K$. In deriving
the above expression, we have used the U$(1)$$\times Z_{2}$ symmetry
of the XXZ chain, which implies that $\langle T_{\tau}\left[S^{+}\left(\mathbf{r}_{1}\right)S^{+}\left(\mathbf{r}_{2}\right)\right]\rangle_{0}=\langle T_{\tau}\left[S^{-}\left(\mathbf{r}_{1}\right)S^{-}\left(\mathbf{r}_{2}\right)\right]\rangle_{0}=0$,
etc. Thus, even if the spin-response of the metallic host lacks the
in-plane spin-rotation {[}U$(1)${]} symmetry of the spin-chain due
to, e.g. spin-orbit interactions (i.e. $\chi^{xy}\left(\mathbf{R},\tau\right)\neq0$,
etc.), the leading corrections to the free-energy are insensitive
to it. However, one factor that complicates the weak-coupling analysis
is the fact that the spin correlation functions of the host electrons
have very different behavior as compared to the spin correlators of
the uncoupled spin chain. In particular, the conformal invariance
(e.g., homogeneous scaling upon space and time rescaling) present
in the XY model, Eq.~\eqref{eq:H_XXZ_Bosonized}), implies the spin
correlators of the uncoupled chain read:\cite{giamarchi_book_1d}\begin{widetext}
\begin{align}
\left\langle T_{\tau}S^{z}\left(\mathbf{r}\right)S^{z}\left(0\right)\right\rangle _{0} & =-\frac{K}{4\pi^{2}}\left[\frac{\left(\frac{\pi T}{u}\right)^{2}}{\sinh^{2}\left(\frac{\pi T\left(x+iu\tau\right)}{u}\right)}+\text{h.c.}\right]+\frac{\cos\left(\frac{\pi x}{a_{0}}\right)}{a_{0}^{2}\pi}\frac{\left(\frac{\pi Ta_{0}}{u}\right)^{2K}}{\sinh^{K}\left(\frac{\pi T\left(x+iu\tau\right)}{u}\right)\sinh^{K}\left(\frac{\pi T\left(x-iu\tau\right)}{u}\right)},\label{eq:SzSz}\\
\left\langle T_{\tau}S^{+}\left(\mathbf{r}\right)S^{-}\left(0\right)\right\rangle _{0} & =\frac{\cos\left(\frac{\pi x}{a_{0}}\right)}{2\pi a_{0}^{2}}\frac{\left(\frac{\pi Ta_{0}}{u}\right)^{1/2K}}{\sinh^{\frac{1}{4K}}\left(\frac{\pi T\left(x+iu\tau\right)}{u}\right)\sinh^{\frac{1}{4K}}\left(\frac{\pi T\left(x-iu\tau\right)}{u}\right)},\label{eq:S+S-}
\end{align}
 \end{widetext} where we have kept the leading terms at large distances
and times. The uniform component of $S_{j}^{+}$ {[}$\sim e^{-i\Theta\left(x\right)}\cos2\Phi\left(x\right)${]}
is less relevant than the staggered part {[}$\sim e^{ix\pi/a_{0}}e^{-i\Theta\left(x_{j}\right)}${]}
and will be neglected in what follows.

Estimating the temperature dependence of the various contributions
to Eq.~\eqref{eq:Delta_F} is possible analyzing the spin-response
function in the electron gas at long times and distances. In Fourier
representation, this reduces to studying the polarization function
$\chi^{ab}\left(\mathbf{Q},\omega_{n}\right)$ near the point$\left(\mathbf{Q},\omega_{n}\right)=0$,
with $\mathbf{Q}=\left(Q_{x},Q_{y},Q_{z}\right)$ the 3D-wavevector
in the metal. As is well-known,\cite{mahan2000} this point is singular
and consequently the limits $\mathbf{Q}\rightarrow0,\omega_{n}\rightarrow0$
do not commute, despite the fact that, in the present case, the singularity
is expected to be smoothen by an integration over the momenta perpendicular
to the spin-chain. Consequently, we only need to retain the wavevector
$Q_{x}$ parallel to the spin-chain. Let us first study the response
in the regime $Q_{x}\neq0,\omega_{n}\rightarrow0$. From general considerations,
for a normal Fermi liquid, the long time dynamics in this regime is
dominated by particle-hole excitations, which yield an Ohmic behavior

\begin{align}
\chi_{x}^{ab}\left(Q_{x},\omega_{n}\right) & \propto f\left(Q_{x}\right)\left|\omega_{n}\right|,\label{eq:}
\end{align}
 for a wide range of momentum transfer $Q_{x}<2k_{F}$. Here we have
defined $\chi_{x}^{ab}\left(Q_{x},\omega_{n}\right)\equiv\int dQ_{y}dQ_{z}\ \chi^{ab}\left(\mathbf{Q},\omega_{n}\right)$,
and $f\left(Q_{x}\right)$ is a smooth function of $Q_{x}$ presenting
no singularities. Thus, after Fourier transformation one obtains 
\begin{equation}
\chi_{x}^{ab}\left(Q_{x},\tau\gg\omega_{c}^{-1}\right)\propto f\left(Q_{x}\right)\frac{T^{2}}{\sin^{2}\left(\pi T\tau\right)},\label{eq:ohmic_response}
\end{equation}
 with $\omega_{c}\simeq E_{F}$. The lattice parameter of the spin-chain
$a_{0}$ determines a characteristic wavevector $Q_{x}\simeq a_{0}^{-1}$,
and the result in Eq. (\ref{eq:ohmic_response}) can be interpreted
as the Ohmic response arising from a slab of metal of width $\sim a_{0}$,
constituting the ``local'' environment seen by each spin in the
chain. This behavior has to be compared with the response in the opposite
regime $\omega_{n}\neq0,Q_{x}\rightarrow0$ 
\begin{align}
\chi_{x}^{ab}\left(Q_{x},\omega_{n}\right) & \propto g\left(\omega_{n}\right)Q_{x}^{2},\label{eq:-1}
\end{align}
 which for an electron gas in the diffusive limit\cite{akkermans}
results in the stronger decay with distance 
\begin{equation}
\chi_{x}^{ab}\left(x\gg a_{0},\omega_{n}\right)\propto\frac{g\left(\omega_{n}\right)}{x^{3}}.\label{eq:non_local_response}
\end{equation}
 Based on these qualitative arguments, we conclude that the most relevant
contribution to $\Delta F$ arises from time decay in Eq. (\ref{eq:ohmic_response}),
and we neglect the more irrelevant effects coming from Eq. (\ref{eq:non_local_response}).

Using these approximations in Eq.~\eqref{eq:Delta_F}) yields 
\begin{align}
\frac{\Delta F}{L} & =g_{z,u}^{2}\frac{KA_{u}^{z}}{2^{6}\pi^{2}u^{2}k_{F}}T^{3}\nonumber \\
 & +g_{z,s}^{2}\frac{A_{s}^{z}}{2^{4}k_{F}a_{0}^{2}\pi^{3}}\left(\frac{\pi a_{0}}{u}\right)^{2K}T^{1+2K}\nonumber \\
 & +g_{\perp,s}^{2}\frac{A_{s}^{\perp}}{2^{5}k_{F}a_{0}^{2}\pi^{3}}\left(\frac{\pi a_{0}}{u}\right)^{1/2K}T^{1+1/2K},
\end{align}
 where we have defined the dimensionless couplings $g_{z,u}\equiv J_{K}^{z}/v_{F}k_{F}$,
$g_{z,s}\equiv J_{K}^{z}/v_{F}k_{F}$ and $g_{\perp,s}\equiv J_{K}^{\perp}/v_{F}k_{F},$
and where $A_{u}^{z}$ , $A_{s}^{z}$ and $A_{s}^{\perp}$ are non-universal
numerical coefficients.%

\section{\label{sec:spin_correlation_function}Spin-spin correlation functions}

Starting from the original spin representation (i.e., before the rotation
$\mathcal{U}$), the transverse spin-correlator is defined as 
\begin{align}
\mathcal{C}_{n}^{+-}\left(\tau\right) & \equiv\left\langle T_{\tau}S_{i+n}^{+}\left(\tau\right)S_{i}^{-}\left(0\right)\right\rangle _{\mathcal{H}},\\
 & =\frac{\int\mathcal{D}\left[S\right]\; e^{-\mathcal{S}}e^{\mathcal{H}\tau}S_{i+n}^{+}e^{-\mathcal{H}\tau}S_{i}^{-}}{\int\mathcal{D}\left[S\right]\; e^{-\mathcal{S}}},
\end{align}
 where the average is taken with respect to the total Hamiltonian
of the system Eq. (\ref{eq:H_total}). The correlation function is
a physical quantity that does not depend on our particular choice
of representation. Using the transformation $\mathcal{U}$, the Hamiltonian
is transformed as in Eqs. \ref{eq:H_F_transformed} and \ref{eq:H_K_transformed},
while the spin operators become $\mathcal{U}^{\dagger}S_{i}^{+}\mathcal{U}=S_{i}^{+}e^{i\gamma\phi_{i,\sigma}^{R}\left(0\right)}$
(see Eq. \ref{eq:transformation_2}). Thus, 
\begin{align}
\mathcal{C}_{n}^{+-}\left(\tau\right) & =\langle T_{\tau}\left(\mathcal{U}^{\dagger}e^{\mathcal{H}\tau}\mathcal{U}\right)\left(\mathcal{U}^{\dagger}S_{i+n}^{+}\mathcal{U}\right)\nonumber \\
 & \qquad\times\left(\mathcal{U}^{\dagger}e^{-\mathcal{H}\tau}\mathcal{U}\right)\left(\mathcal{U}^{\dagger}S_{i}^{-}\mathcal{U}\right)\rangle\\
 & =\left\langle T_{\tau}e^{\tilde{\mathcal{H}}\tau}S_{i+n}^{+}e^{i\frac{\gamma}{\sqrt{2}}\varphi_{i+n}}e^{-\tilde{\mathcal{H}}\tau}S_{i}^{-}e^{-i\frac{\gamma}{\sqrt{2}}\varphi_{i}}\right\rangle ,\\
 & =\left\langle T_{\tau}S_{i+n}^{+}\left(\tau\right)e^{i\frac{\gamma}{\sqrt{2}}\varphi_{i+n}\left(\tau\right)}S_{i}^{-}\left(0\right)e^{-i\frac{\gamma}{\sqrt{2}}\varphi_{i}\left(0\right)}\right\rangle _{\tilde{\mathcal{H}}}.
\end{align}

Imposing $\gamma=\sqrt{2}$, and near the strong-coupling Kondo fixed-point,
the Hamiltonian $\tilde{\mathcal{H}}$ maps onto Eqs. (\ref{eq:H_K_projected})
and (\ref{eq:H_xxz_projected}) (the conduction electron term $\mathcal{H}_{F}$
is not changed since it doesn't depend on spin operators). We therefore
eliminate the spin-degrees of freedom in the correlation function
and obtain

\begin{align}
\mathcal{C}_{n}^{+-}\left(\tau\right) & =\left\langle T_{\tau}e^{i\varphi_{i+n}\left(\tau\right)}e^{-i\varphi_{i}\left(0\right)}\right\rangle _{\tilde{\mathcal{H}}^{\prime}},\label{eq:correlator_generic}\\
 & =\frac{\int\mathcal{D}\left[\varphi\right]\; e^{-\mathcal{S}^{\prime}\left[\varphi\right]}e^{i\varphi_{i+n}\left(\tau\right)}e^{-i\varphi_{i}\left(0\right)}}{\int\mathcal{D}\left[\varphi\right]\; e^{-\mathcal{S}^{\prime}\left[\varphi\right]}},
\end{align}
 where the averages are now taken with respect to the effective action
Eq. (\ref{eq:action_strong_coupling}).

\subsection{\label{sub:spin_correlator_Kondo}Spin-correlators in the disordered
Kondo-singlet phase}

In the disordered ``Kondo'' phase $J_{K}^{\perp}\gg\left\{ J_{H}^{\perp},J_{H}^{z}\right\} $,
the Heisenberg coupling $J_{H}^{\perp}$ in $\mathcal{S}_{H}^{\prime}$
Eq. (\ref{eq:S_H}) is a suitable expansion parameter to compute the
correlation function. Explicitly\begin{widetext}

\begin{align}
\mathcal{C}_{n}^{+-}\left(\tau\right) & =\frac{\int\mathcal{D}\left[\varphi\right]\; e^{-\mathcal{S}_{0}^{\prime}\left[\varphi\right]-\mathcal{S}_{H}^{\prime}\left[\varphi\right]}e^{i\varphi_{i+n}\left(\tau\right)}e^{-i\varphi_{i}\left(0\right)}}{\int\mathcal{D}\left[\varphi\right]\; e^{-\mathcal{S}_{0}^{\prime}\left[\varphi\right]-\mathcal{S}_{H}^{\prime}\left[\varphi\right]}},\\
 & =\frac{\int\mathcal{D}\left[\varphi\right]\; e^{-\mathcal{S}_{0}}\left\{ \sum_{m=0}\frac{1}{m!}\prod_{j=1}^{m}\left[\int_{0}^{\beta}d\tau_{j}\;\frac{J_{H}^{\perp}}{4^{2}}\left(\sum_{l}e^{i\varphi_{l+1}\left(\tau_{j}\right)-i\varphi_{l}\left(\tau_{j}\right)}+\text{h.c.}\right)\right]\right\} e^{i\varphi_{i+n}\left(\tau\right)-i\varphi_{i}\left(0\right)}}{\int\mathcal{D}\left[\varphi\right]\; e^{-\mathcal{S}_{0}\left[\varphi\right]}\left\{ \sum_{m=0}\frac{1}{m!}\prod_{j=1}^{m}\left[\int_{0}^{\beta}d\tau_{j}\;\frac{J_{H}^{\perp}}{4^{2}}\left(\sum_{l}e^{i\varphi_{l+1}\left(\tau_{j}\right)-i\varphi_{l}\left(\tau_{j}\right)}+\text{h.c.}\right)\right]\right\} },\\
 & \simeq\left(\frac{J_{H}^{\perp}}{4^{2}}\right)^{n}\frac{1}{n!}\int_{0}^{\beta}d\tau_{1}d\tau_{2}\dots d\tau_{n}\;\left\langle T_{\tau}e^{i\varphi_{i+n}\left(\tau\right)}e^{-i\varphi_{i+n}\left(\tau_{n}\right)}e^{i\varphi_{i+n-1}\left(\tau_{n}\right)}\dots e^{-i\varphi_{i+1}\left(\tau_{1}\right)}e^{i\varphi_{i}\left(\tau_{1}\right)}e^{-i\varphi_{i}\left(0\right)}\right\rangle _{0},
\end{align}
 \end{widetext} where we have truncated the perturbative expansion
at leading order. This is to be expected, since the Heisenberg term
only couples nearest-neighbors, and therefore spins at a distance
of $n$ sites only become correlated at order $\left(J_{H}^{\perp}\right)^{n}$
in the perturbative expansion. The product $\left\langle T_{\tau}e^{i\varphi_{i+n}\left(\tau\right)}e^{-i\varphi_{i+n}\left(\tau_{n}\right)}\dots e^{-i\varphi_{i+1}\left(\tau_{1}\right)}e^{i\varphi_{i}\left(\tau_{1}\right)}e^{-i\varphi_{i}\left(0\right)}\right\rangle _{\mathcal{S}_{0}^{\prime}}$
can be calculated using Wick's theorem. For compactness in notation,
we define the (local) two-point correlation function 
\begin{align}
F\left(\tau\right) & \equiv\left\langle T_{\tau}e^{i\varphi_{m}\left(\tau_{1}\right)}e^{-i\varphi_{m}\left(\tau_{2}\right)}\right\rangle _{0},\label{eq:F_def}\\
 & \simeq\frac{1}{\left(\frac{E_{0}\tau}{\sqrt{a}}\right)^{2}+1} & \left(\text{at }T=0\right),\label{eq:F_explicit_form}
\end{align}
 where $a$ is a numerical factor $a=4e^{-2\gamma_{E}}\simeq1.261...$,
and its Fourier transform 
\begin{align}
F\left(\omega_{m}\right) & \equiv\int_{0}^{\beta}d\tau\; e^{i\omega_{m}\tau}F\left(\tau\right),\nonumber \\
 & =\frac{\pi e^{-\gamma_{E}}}{E_{0}}\exp\left[-\frac{\left|\omega_{m}\right|2e^{-\gamma_{E}}}{E_{0}}\right].\label{eq:F_Fourier}
\end{align}
 Then, the expression for $\mathcal{C}_{n}^{+-}\left(\tau\right)$
compactly writes 
\begin{align}
\mathcal{C}_{n}^{+-}\left(\tau\right) & =\begin{cases}
\left(\frac{J_{H}^{\perp}}{2^{2}}\right)^{n}\frac{1}{\beta}\sum_{\omega_{m}}e^{i\omega_{m}\tau}\left[F\left(\omega_{m}\right)\right]^{n+1}, & \quad n>0\\
F\left(\tau\right). & \quad n=0
\end{cases}\label{eq:C_Kondo_limit}
\end{align}
 One particularly interesting case is the local dynamical correlation
$\mathcal{C}_{0}^{+-}\left(\tau\right)\propto\tau^{-2}$ {[}cf. Eq.
\ref{eq:F_explicit_form}{]}, which encodes the properties of a local
Fermi liquid. Another one is the static, non-local correlation 
\begin{align}
\mathcal{C}_{n}^{+-}\left(0\right) & =\frac{1}{2}\frac{e^{-n/\xi_{c}}}{n+1},\label{eq:C_non_local_static}
\end{align}
 where we have defined the correlation length $\xi_{c}\equiv1/\ln\left|\frac{8e^{\gamma_{E}}}{\pi}\frac{E_{0}}{J_{H}^{\perp}}\right|$.

\subsection{\label{sub:spin_correlator_dissipative}Spin-correlations in the
ordered phase}

In the ordered phase, the dynamics of the spin-chain is effectively
given by the action of the Goldstone mode Eqs. (\ref{eq:S_Goldstone_mode}).
With this Gaussian action, and using the saddle-point equations in
Eq. (\ref{eq:S_psi}) to express $e^{i\varphi_{i}\left(\tau\right)}=\psi_{i}^{*}\left(\tau\right)$,
we can calculate the spin-correlation in Eq. (\ref{eq:correlator_generic})
as 
\begin{align}
\mathcal{C}_{n}^{+-}\left(\tau\right) & =\left\langle T_{\tau}e^{i\varphi_{i+n}\left(\tau\right)}e^{-i\varphi_{i}\left(0\right)}\right\rangle ,\\
 & =\left(-1\right)^{n}\psi_{0}^{2}\left\langle T_{\tau}e^{-i\vartheta_{i+n}\left(\tau\right)}e^{i\vartheta_{i}\left(0\right)}\right\rangle ,\\
 & =\left(-1\right)^{n}\psi_{0}^{2}e^{-\frac{1}{2}\left\langle T_{\tau}\left[\vartheta_{i+n}\left(\tau\right)-\vartheta_{i}\left(0\right)\right]^{2}\right\rangle },
\end{align}
 where the factor $\left(-1\right)^{n}$ comes from the antiferromagnetic
correlations induced by the coupling $J_{H}^{\perp}>0$, \cite{Lobos12_comment_sign_JH}
and where 
\begin{align}
\left\langle T_{\tau}\left[\vartheta_{i+n}\left(\tau\right)-\vartheta_{i}\left(0\right)\right]^{2}\right\rangle  & =\frac{1}{\beta L}\sum_{k,\omega_{m}}\frac{1-e^{ika_{0}n-i\omega_{m}\tau}}{\frac{\pi}{e^{2\gamma}}\frac{\psi_{0}^{2}\left|\omega_{m}\right|}{E_{0}^{2}}+\frac{16\psi_{0}^{2}k^{2}}{J_{H}^{\perp}}}.
\end{align}
 We study this correlation function in two different limits: the local
limit $n=0$ , and $\tau=0$. In the first case we have

\begin{align}
 & \left\langle T_{\tau}\left[\vartheta_{i}\left(\tau\right)-\vartheta_{i}\left(0\right)\right]^{2}\right\rangle =\nonumber \\
 & =\frac{1}{\beta L}\sum_{k,\omega_{m}}\frac{1-e^{-i\omega_{m}\tau}}{\frac{\pi}{e^{2\gamma}}\frac{\psi_{0}^{2}\left|\omega_{m}\right|}{E_{0}^{2}}+\frac{16\psi_{0}^{2}k^{2}}{J_{H}^{\perp}}},\label{eq:-2}\\
 & =\left(\frac{1}{8\pi}\right)^{2}\frac{J_{H}^{\perp}}{\psi_{0}^{2}}\int d\omega dk\frac{1-e^{-i\omega\tau}}{\frac{\pi}{16e^{2\gamma}}\frac{J_{H}^{\perp}\left|\omega\right|}{E_{0}^{2}}+k^{2}},\label{eq:-3}\\
 & \simeq\left(\frac{1}{2\pi}\right)^{2}\frac{\pi e^{\gamma}\left(J_{H}^{\perp}\right)^{3/2}E_{0}}{\sqrt{2}8\psi_{0}^{2}}\frac{1}{\sqrt{\tau}},\label{eq:-4}
\end{align}
 where we have introduced the short-time cutoff $\tau_{0}$. With
similar tools it can be shown that the static correlation decays as
$\left\langle T_{\tau}\left[\vartheta_{i+n}\left(0\right)-\vartheta_{i}\left(0\right)\right]^{2}\right\rangle \sim\left|n\right|^{-1}$.

\bibliographystyle{apsrev}

\begin{thebibliography}{79}
\expandafter\ifx\csname natexlab\endcsname\relax\def\natexlab#1{#1}\fi
\expandafter\ifx\csname bibnamefont\endcsname\relax
  \def\bibnamefont#1{#1}\fi
\expandafter\ifx\csname bibfnamefont\endcsname\relax
  \def\bibfnamefont#1{#1}\fi
\expandafter\ifx\csname citenamefont\endcsname\relax
  \def\citenamefont#1{#1}\fi
\expandafter\ifx\csname url\endcsname\relax
  \def\url#1{\texttt{#1}}\fi
\expandafter\ifx\csname urlprefix\endcsname\relax\def\urlprefix{URL }\fi
\providecommand{\bibinfo}[2]{#2}
\providecommand{\eprint}[2][]{\url{#2}}

\bibitem[{\citenamefont{Wiesendanger}(2009)}]{Wiesendanger09_Spins_on_surfaces_review}
\bibinfo{author}{\bibfnamefont{R.}~\bibnamefont{Wiesendanger}},
  \bibinfo{journal}{Rev. Mod. Phys.} \textbf{\bibinfo{volume}{81}},
  \bibinfo{pages}{1495} (\bibinfo{year}{2009}).

\bibitem[{\citenamefont{Jamneala et~al.}(2001)\citenamefont{Jamneala, Madhavan,
  and Crommie}}]{jamneala01}
\bibinfo{author}{\bibfnamefont{T.}~\bibnamefont{Jamneala}},
  \bibinfo{author}{\bibfnamefont{V.}~\bibnamefont{Madhavan}}, \bibnamefont{and}
  \bibinfo{author}{\bibfnamefont{M.~F.} \bibnamefont{Crommie}},
  \bibinfo{journal}{Phys. Rev. Lett.} \textbf{\bibinfo{volume}{87}},
  \bibinfo{pages}{256804} (\bibinfo{year}{2001}).

\bibitem[{\citenamefont{Hirjibehedin et~al.}(2006)\citenamefont{Hirjibehedin,
  Lutz, and Heinrich}}]{Hirjibehedin06_Mn_spin_chains}
\bibinfo{author}{\bibfnamefont{C.~F.} \bibnamefont{Hirjibehedin}},
  \bibinfo{author}{\bibfnamefont{C.~P.} \bibnamefont{Lutz}}, \bibnamefont{and}
  \bibinfo{author}{\bibfnamefont{A.~J.} \bibnamefont{Heinrich}},
  \bibinfo{journal}{Science} \textbf{\bibinfo{volume}{312}},
  \bibinfo{pages}{1021} (\bibinfo{year}{2006}).

\bibitem[{\citenamefont{Zhou et~al.}(2010)\citenamefont{Zhou, Wiebe, Lounis,
  Vedmedenko, Meier, Bl{\"u}gel, Dederichs, and
  Wiesendanger}}]{Zhou10_RKKY_on_atomic_scale}
\bibinfo{author}{\bibfnamefont{L.}~\bibnamefont{Zhou}},
  \bibinfo{author}{\bibfnamefont{J.}~\bibnamefont{Wiebe}},
  \bibinfo{author}{\bibfnamefont{S.}~\bibnamefont{Lounis}},
  \bibinfo{author}{\bibfnamefont{E.}~\bibnamefont{Vedmedenko}},
  \bibinfo{author}{\bibfnamefont{F.}~\bibnamefont{Meier}},
  \bibinfo{author}{\bibfnamefont{S.}~\bibnamefont{Bl{\"u}gel}},
  \bibinfo{author}{\bibfnamefont{P.~H.} \bibnamefont{Dederichs}},
  \bibnamefont{and}
  \bibinfo{author}{\bibfnamefont{R.}~\bibnamefont{Wiesendanger}},
  \bibinfo{journal}{Nature Physics} \textbf{\bibinfo{volume}{6}},
  \bibinfo{pages}{187} (\bibinfo{year}{2010}).

\bibitem[{\citenamefont{Serrate et~al.}(2010)\citenamefont{Serrate, Ferriani,
  Yoshida, Hla, Menzel, von Bergmann, Heinze, and
  Kubetzka}}]{Serrate10_Imaging_and_manipulating_spin_of_individual_atoms}
\bibinfo{author}{\bibfnamefont{D.}~\bibnamefont{Serrate}},
  \bibinfo{author}{\bibfnamefont{P.}~\bibnamefont{Ferriani}},
  \bibinfo{author}{\bibfnamefont{Y.}~\bibnamefont{Yoshida}},
  \bibinfo{author}{\bibfnamefont{S.-W.} \bibnamefont{Hla}},
  \bibinfo{author}{\bibfnamefont{M.}~\bibnamefont{Menzel}},
  \bibinfo{author}{\bibfnamefont{K.}~\bibnamefont{von Bergmann}},
  \bibinfo{author}{\bibfnamefont{S.}~\bibnamefont{Heinze}}, \bibnamefont{and}
  \bibinfo{author}{\bibfnamefont{R.}~\bibnamefont{Kubetzka},
  \bibfnamefont{Andre~andWiesendanger}}, \bibinfo{journal}{Nature
  Nanotechnology} \textbf{\bibinfo{volume}{5}}, \bibinfo{pages}{350}
  (\bibinfo{year}{2010}).

\bibitem[{\citenamefont{Li et~al.}(1998)\citenamefont{Li, Schneider, Berndt,
  and Delley}}]{Li98_Kondo_effect_on_single_adatoms}
\bibinfo{author}{\bibfnamefont{J.}~\bibnamefont{Li}},
  \bibinfo{author}{\bibfnamefont{W.-D.} \bibnamefont{Schneider}},
  \bibinfo{author}{\bibfnamefont{R.}~\bibnamefont{Berndt}}, \bibnamefont{and}
  \bibinfo{author}{\bibfnamefont{B.}~\bibnamefont{Delley}},
  \bibinfo{journal}{Phys. Rev. Lett.} \textbf{\bibinfo{volume}{80}},
  \bibinfo{pages}{2893} (\bibinfo{year}{1998}).

\bibitem[{\citenamefont{Madhavan et~al.}(1998)\citenamefont{Madhavan, Chen,
  Jamneala, Crommie, and
  Wingreen}}]{Madhavan98_Tunneling_into_single_Kondo_adatom}
\bibinfo{author}{\bibfnamefont{V.}~\bibnamefont{Madhavan}},
  \bibinfo{author}{\bibfnamefont{W.}~\bibnamefont{Chen}},
  \bibinfo{author}{\bibfnamefont{T.}~\bibnamefont{Jamneala}},
  \bibinfo{author}{\bibfnamefont{M.~F.} \bibnamefont{Crommie}},
  \bibnamefont{and} \bibinfo{author}{\bibfnamefont{N.~S.}
  \bibnamefont{Wingreen}}, \bibinfo{journal}{Science}
  \textbf{\bibinfo{volume}{280}}, \bibinfo{pages}{567} (\bibinfo{year}{1998}).

\bibitem[{\citenamefont{Knorr et~al.}(2002)\citenamefont{Knorr, Schneider,
  Diekhoner, Wahl, and Kern}}]{knorr02}
\bibinfo{author}{\bibfnamefont{N.}~\bibnamefont{Knorr}},
  \bibinfo{author}{\bibfnamefont{M.~A.} \bibnamefont{Schneider}},
  \bibinfo{author}{\bibfnamefont{L.}~\bibnamefont{Diekhoner}},
  \bibinfo{author}{\bibfnamefont{P.}~\bibnamefont{Wahl}}, \bibnamefont{and}
  \bibinfo{author}{\bibfnamefont{K.}~\bibnamefont{Kern}},
  \bibinfo{journal}{Phys. Rev. Lett.} \textbf{\bibinfo{volume}{88}},
  \bibinfo{pages}{096804} (\bibinfo{year}{2002}).

\bibitem[{\citenamefont{Hewson}(1993)}]{hewson}
\bibinfo{author}{\bibfnamefont{A.~C.} \bibnamefont{Hewson}},
  \emph{\bibinfo{title}{The Kondo Problem to Heavy Fermions}}
  (\bibinfo{publisher}{Cambridge University Press},
  \bibinfo{address}{Cambridge}, \bibinfo{year}{1993}).

\bibitem[{\citenamefont{Ruderman and Kittel}(1954)}]{ruderman54}
\bibinfo{author}{\bibfnamefont{M.~A.} \bibnamefont{Ruderman}} \bibnamefont{and}
  \bibinfo{author}{\bibfnamefont{C.}~\bibnamefont{Kittel}},
  \bibinfo{journal}{Phys. Rev.} \textbf{\bibinfo{volume}{96}},
  \bibinfo{pages}{66} (\bibinfo{year}{1954}).

\bibitem[{\citenamefont{L{\"o}hneysen}(1999)}]{Lohneysen99_Ce_review}
\bibinfo{author}{\bibfnamefont{H.}~\bibnamefont{L{\"o}hneysen}},
  \bibinfo{journal}{J. Magn. Magn. Mat.} \textbf{\bibinfo{volume}{200}},
  \bibinfo{pages}{532} (\bibinfo{year}{1999}).

\bibitem[{\citenamefont{Baibich et~al.}(1988)\citenamefont{Baibich, Broto,
  Fert, NguyenVanDau, Petroff, Etienne, Creuzet, Friederich, and
  Chazelas}}]{Baibich88_GMR_in_layered_magnetic_superlattices}
\bibinfo{author}{\bibfnamefont{M.~N.} \bibnamefont{Baibich}},
  \bibinfo{author}{\bibfnamefont{J.~M.} \bibnamefont{Broto}},
  \bibinfo{author}{\bibfnamefont{A.}~\bibnamefont{Fert}},
  \bibinfo{author}{\bibfnamefont{F.}~\bibnamefont{NguyenVanDau}},
  \bibinfo{author}{\bibfnamefont{F.}~\bibnamefont{Petroff}},
  \bibinfo{author}{\bibfnamefont{P.}~\bibnamefont{Etienne}},
  \bibinfo{author}{\bibfnamefont{G.}~\bibnamefont{Creuzet}},
  \bibinfo{author}{\bibfnamefont{A.}~\bibnamefont{Friederich}},
  \bibnamefont{and} \bibinfo{author}{\bibfnamefont{J.}~\bibnamefont{Chazelas}},
  \bibinfo{journal}{Phys. Rev. Lett.} \textbf{\bibinfo{volume}{61}},
  \bibinfo{pages}{2472} (\bibinfo{year}{1988}).

\bibitem[{\citenamefont{Wahl et~al.}(2007)\citenamefont{Wahl, Simon,
  Diekh\"oner, Stepanyuk, Bruno, Schneider, and
  Kern}}]{Wahl07_Exchange_Interaction_between_Single_Magnetic_Adatoms}
\bibinfo{author}{\bibfnamefont{P.}~\bibnamefont{Wahl}},
  \bibinfo{author}{\bibfnamefont{P.}~\bibnamefont{Simon}},
  \bibinfo{author}{\bibfnamefont{L.}~\bibnamefont{Diekh\"oner}},
  \bibinfo{author}{\bibfnamefont{V.~S.} \bibnamefont{Stepanyuk}},
  \bibinfo{author}{\bibfnamefont{P.}~\bibnamefont{Bruno}},
  \bibinfo{author}{\bibfnamefont{M.~A.} \bibnamefont{Schneider}},
  \bibnamefont{and} \bibinfo{author}{\bibfnamefont{K.}~\bibnamefont{Kern}},
  \bibinfo{journal}{Phys. Rev. Lett.} \textbf{\bibinfo{volume}{98}},
  \bibinfo{pages}{056601} (\bibinfo{year}{2007}).

\bibitem[{\citenamefont{Loth et~al.}(2012)\citenamefont{Loth, Baumann, Lutz,
  Eigler, and Heinrich}}]{Loth12_Bistability_in_atomic_scale_AFM}
\bibinfo{author}{\bibfnamefont{S.}~\bibnamefont{Loth}},
  \bibinfo{author}{\bibfnamefont{S.}~\bibnamefont{Baumann}},
  \bibinfo{author}{\bibfnamefont{C.~P.} \bibnamefont{Lutz}},
  \bibinfo{author}{\bibfnamefont{D.~M.} \bibnamefont{Eigler}},
  \bibnamefont{and} \bibinfo{author}{\bibfnamefont{A.~J.}
  \bibnamefont{Heinrich}}, \bibinfo{journal}{Science}
  \textbf{\bibinfo{volume}{335}}, \bibinfo{pages}{196} (\bibinfo{year}{2012}).

\bibitem[{\citenamefont{Giamarchi}(2004)}]{giamarchi_book_1d}
\bibinfo{author}{\bibfnamefont{T.}~\bibnamefont{Giamarchi}},
  \emph{\bibinfo{title}{Quantum Physics in One Dimension}}
  (\bibinfo{publisher}{Oxford University Press}, \bibinfo{address}{Oxford},
  \bibinfo{year}{2004}).

\bibitem[{\citenamefont{Caldeira and Leggett}(1981)}]{caldeira&leggett81}
\bibinfo{author}{\bibfnamefont{A.~O.} \bibnamefont{Caldeira}} \bibnamefont{and}
  \bibinfo{author}{\bibfnamefont{A.~J.} \bibnamefont{Leggett}},
  \bibinfo{journal}{Phys. Rev. Lett.} \textbf{\bibinfo{volume}{46}},
  \bibinfo{pages}{211} (\bibinfo{year}{1981}).

\bibitem[{\citenamefont{Sch{\"o}n and
  Zaikin}(1990)}]{schoen90_review_ultrasmall_tunnel_junctions}
\bibinfo{author}{\bibfnamefont{G.}~\bibnamefont{Sch{\"o}n}} \bibnamefont{and}
  \bibinfo{author}{\bibfnamefont{A.~D.} \bibnamefont{Zaikin}},
  \bibinfo{journal}{Physics Reports} \textbf{\bibinfo{volume}{198}},
  \bibinfo{pages}{237 } (\bibinfo{year}{1990}), ISSN \bibinfo{issn}{0370-1573}.

\bibitem[{\citenamefont{Fazio and van~der
  Zant}(2001)}]{fazio01_review_superconducting_networks}
\bibinfo{author}{\bibfnamefont{R.}~\bibnamefont{Fazio}} \bibnamefont{and}
  \bibinfo{author}{\bibfnamefont{H.}~\bibnamefont{van~der Zant}},
  \bibinfo{journal}{Physics Reports} \textbf{\bibinfo{volume}{355}},
  \bibinfo{pages}{235} (\bibinfo{year}{2001}).

\bibitem[{\citenamefont{Tewari and
  Toner}(2006)}]{Tewari06_Dissipate_locally_couple_globally}
\bibinfo{author}{\bibfnamefont{S.}~\bibnamefont{Tewari}} \bibnamefont{and}
  \bibinfo{author}{\bibfnamefont{J.}~\bibnamefont{Toner}},
  \bibinfo{journal}{Europhysics Letters} \textbf{\bibinfo{volume}{74}}
  (\bibinfo{year}{2006}).

\bibitem[{\citenamefont{Goswami and
  Chakravarty}(2006)}]{goswami06_josephson_array}
\bibinfo{author}{\bibfnamefont{P.}~\bibnamefont{Goswami}} \bibnamefont{and}
  \bibinfo{author}{\bibfnamefont{S.}~\bibnamefont{Chakravarty}},
  \bibinfo{journal}{Phys. Rev. B} \textbf{\bibinfo{volume}{73}},
  \bibinfo{pages}{094516} (\bibinfo{year}{2006}).

\bibitem[{\citenamefont{Refael et~al.}(2007)\citenamefont{Refael, Demler, Oreg,
  and Fisher}}]{refael07_SN_transition_in_grains&nanowires}
\bibinfo{author}{\bibfnamefont{G.}~\bibnamefont{Refael}},
  \bibinfo{author}{\bibfnamefont{E.}~\bibnamefont{Demler}},
  \bibinfo{author}{\bibfnamefont{Y.}~\bibnamefont{Oreg}}, \bibnamefont{and}
  \bibinfo{author}{\bibfnamefont{D.~S.} \bibnamefont{Fisher}},
  \bibinfo{journal}{Phys. Rev. B} \textbf{\bibinfo{volume}{75}},
  \bibinfo{pages}{014522} (\bibinfo{year}{2007}).

\bibitem[{\citenamefont{Lobos and
  Giamarchi}(2011)}]{Lobos11_sit_in_dissipative_jjas}
\bibinfo{author}{\bibfnamefont{A.~M.} \bibnamefont{Lobos}} \bibnamefont{and}
  \bibinfo{author}{\bibfnamefont{T.}~\bibnamefont{Giamarchi}},
  \bibinfo{journal}{Phys. Rev. B} \textbf{\bibinfo{volume}{84}},
  \bibinfo{pages}{024523} (\bibinfo{year}{2011}).

\bibitem[{\citenamefont{Castro~Neto et~al.}(1997)\citenamefont{Castro~Neto,
  Chamon, and Nayak}}]{castroneto97_open_luttinger_liquids}
\bibinfo{author}{\bibfnamefont{A.~H.} \bibnamefont{Castro~Neto}},
  \bibinfo{author}{\bibfnamefont{C.}~\bibnamefont{Chamon}}, \bibnamefont{and}
  \bibinfo{author}{\bibfnamefont{C.}~\bibnamefont{Nayak}},
  \bibinfo{journal}{Phys. Rev. Lett.} \textbf{\bibinfo{volume}{79}},
  \bibinfo{pages}{4629} (\bibinfo{year}{1997}).

\bibitem[{\citenamefont{Cazalilla et~al.}(2006)\citenamefont{Cazalilla, Sols,
  and Guinea}}]{cazalilla06_dissipative_transition}
\bibinfo{author}{\bibfnamefont{M.~A.} \bibnamefont{Cazalilla}},
  \bibinfo{author}{\bibfnamefont{F.}~\bibnamefont{Sols}}, \bibnamefont{and}
  \bibinfo{author}{\bibfnamefont{F.}~\bibnamefont{Guinea}},
  \bibinfo{journal}{Phys. Rev. Lett.} \textbf{\bibinfo{volume}{97}},
  \bibinfo{pages}{076401} (\bibinfo{year}{2006}).

\bibitem[{\citenamefont{Artemenko and
  Nattermann}(2007)}]{artemenko07_longrange_order_in_1d}
\bibinfo{author}{\bibfnamefont{S.~N.} \bibnamefont{Artemenko}}
  \bibnamefont{and}
  \bibinfo{author}{\bibfnamefont{T.}~\bibnamefont{Nattermann}},
  \bibinfo{journal}{Phys. Rev. Lett.} \textbf{\bibinfo{volume}{99}},
  \bibinfo{pages}{256401} (\bibinfo{year}{2007}).

\bibitem[{\citenamefont{Lobos et~al.}(2009)\citenamefont{Lobos, Iucci,
  M{\"u}ller, and Giamarchi}}]{lobos09_dissipation_scwires}
\bibinfo{author}{\bibfnamefont{A.~M.} \bibnamefont{Lobos}},
  \bibinfo{author}{\bibfnamefont{A.}~\bibnamefont{Iucci}},
  \bibinfo{author}{\bibfnamefont{M.}~\bibnamefont{M{\"u}ller}},
  \bibnamefont{and}
  \bibinfo{author}{\bibfnamefont{T.}~\bibnamefont{Giamarchi}},
  \bibinfo{journal}{Phys. Rev. B} \textbf{\bibinfo{volume}{80}},
  \bibinfo{pages}{214515} (\bibinfo{year}{2009}).

\bibitem[{\citenamefont{Lobos and
  Giamarchi}(2010)}]{Lobos10_Dissipative_phase_fluctuations}
\bibinfo{author}{\bibfnamefont{A.~M.} \bibnamefont{Lobos}} \bibnamefont{and}
  \bibinfo{author}{\bibfnamefont{T.}~\bibnamefont{Giamarchi}},
  \bibinfo{journal}{Phys. Rev. B} \textbf{\bibinfo{volume}{82}},
  \bibinfo{pages}{104517} (\bibinfo{year}{2010}).

\bibitem[{\citenamefont{Orth et~al.}(2008)\citenamefont{Orth, Stanic, and
  Le~Hur}}]{Orth08_Dissipative_quantum_Ising_model_in_cold_atoms}
\bibinfo{author}{\bibfnamefont{P.~P.} \bibnamefont{Orth}},
  \bibinfo{author}{\bibfnamefont{I.}~\bibnamefont{Stanic}}, \bibnamefont{and}
  \bibinfo{author}{\bibfnamefont{K.}~\bibnamefont{Le~Hur}},
  \bibinfo{journal}{Phys. Rev. A} \textbf{\bibinfo{volume}{77}},
  \bibinfo{pages}{051601} (\bibinfo{year}{2008}).

\bibitem[{\citenamefont{Stepanyuk et~al.}(2001)\citenamefont{Stepanyuk,
  Baranov, Bazhanov, Hergert, and
  Katsnelson}}]{Stepanyuk_Magnetic_properties_of_Co_clusters_on_Cu100}
\bibinfo{author}{\bibfnamefont{V.~S.} \bibnamefont{Stepanyuk}},
  \bibinfo{author}{\bibfnamefont{A.~N.} \bibnamefont{Baranov}},
  \bibinfo{author}{\bibfnamefont{D.~I.} \bibnamefont{Bazhanov}},
  \bibinfo{author}{\bibfnamefont{W.}~\bibnamefont{Hergert}}, \bibnamefont{and}
  \bibinfo{author}{\bibfnamefont{A.~A.} \bibnamefont{Katsnelson}},
  \bibinfo{journal}{Surf. Sci.} \textbf{\bibinfo{volume}{482-485}}
  (\bibinfo{year}{2001}).

\bibitem[{\citenamefont{Carbone et~al.}(2011)\citenamefont{Carbone, Gardonio,
  Moras, Lounis, Heide, Bihlmayer, Atodiresei, Dederichs, Bl{\"u}gel, Vlaic
  et~al.}}]{Carbone11_Self-assembled_magnetic_networks_on_surfaces}
\bibinfo{author}{\bibfnamefont{C.}~\bibnamefont{Carbone}},
  \bibinfo{author}{\bibfnamefont{S.}~\bibnamefont{Gardonio}},
  \bibinfo{author}{\bibfnamefont{P.}~\bibnamefont{Moras}},
  \bibinfo{author}{\bibfnamefont{S.}~\bibnamefont{Lounis}},
  \bibinfo{author}{\bibfnamefont{M.}~\bibnamefont{Heide}},
  \bibinfo{author}{\bibfnamefont{G.}~\bibnamefont{Bihlmayer}},
  \bibinfo{author}{\bibfnamefont{N.}~\bibnamefont{Atodiresei}},
  \bibinfo{author}{\bibfnamefont{P.~H.} \bibnamefont{Dederichs}},
  \bibinfo{author}{\bibfnamefont{S.}~\bibnamefont{Bl{\"u}gel}},
  \bibinfo{author}{\bibfnamefont{S.}~\bibnamefont{Vlaic}},
  \bibnamefont{et~al.}, \bibinfo{journal}{Adv. Funct. Mater.}
  \textbf{\bibinfo{volume}{21}}, \bibinfo{pages}{1212} (\bibinfo{year}{2011}).

\bibitem[{\citenamefont{Garate and
  Affleck}(2010)}]{Garate10_Anisotropic_spin_chains}
\bibinfo{author}{\bibfnamefont{I.}~\bibnamefont{Garate}} \bibnamefont{and}
  \bibinfo{author}{\bibfnamefont{I.}~\bibnamefont{Affleck}},
  \bibinfo{journal}{Phys. Rev. B} \textbf{\bibinfo{volume}{81}},
  \bibinfo{pages}{144419} (\bibinfo{year}{2010}).

\bibitem[{\citenamefont{Brezin and Zinn-Justin}(1988)}]{affleck_houches}
\bibinfo{editor}{\bibfnamefont{E.}~\bibnamefont{Brezin}} \bibnamefont{and}
  \bibinfo{editor}{\bibfnamefont{J.}~\bibnamefont{Zinn-Justin}}, eds.,
  \emph{\bibinfo{title}{Fields, Strings and Critical Phenomena}}
  (\bibinfo{publisher}{Elsevier Science Publishers},
  \bibinfo{address}{Amsterdam}, \bibinfo{year}{1988}).

\bibitem[{\citenamefont{Castro~Neto and
  Jones}(2000)}]{CastroNeto00_NFL_in_U_and_Ce_alloys}
\bibinfo{author}{\bibfnamefont{A.~H.} \bibnamefont{Castro~Neto}}
  \bibnamefont{and} \bibinfo{author}{\bibfnamefont{B.~A.} \bibnamefont{Jones}},
  \bibinfo{journal}{Phys. Rev. B} \textbf{\bibinfo{volume}{62}},
  \bibinfo{pages}{14975} (\bibinfo{year}{2000}).

\bibitem[{\citenamefont{Si and
  Steglich}(2010)}]{Si10_Heavy_fermions_and_phase_transitions}
\bibinfo{author}{\bibfnamefont{Q.}~\bibnamefont{Si}} \bibnamefont{and}
  \bibinfo{author}{\bibfnamefont{F.}~\bibnamefont{Steglich}},
  \bibinfo{journal}{Science} \textbf{\bibinfo{volume}{329}}
  (\bibinfo{year}{2010}).

\bibitem[{\citenamefont{Zachar and Tsvelik}(2001)}]{zachar_exotic_kondo}
\bibinfo{author}{\bibfnamefont{O.}~\bibnamefont{Zachar}} \bibnamefont{and}
  \bibinfo{author}{\bibfnamefont{A.~M.} \bibnamefont{Tsvelik}},
  \bibinfo{journal}{Phys. Rev. B} \textbf{\bibinfo{volume}{64}},
  \bibinfo{pages}{033103} (\bibinfo{year}{2001}),
  \bibinfo{note}{cond-mat/9909296}.

\bibitem[{\citenamefont{Zachar}(2001)}]{Zachar01_Staggered_phases_1D_Kondo_Heisenberg_model}
\bibinfo{author}{\bibfnamefont{O.}~\bibnamefont{Zachar}},
  \bibinfo{journal}{Phys. Rev. B} \textbf{\bibinfo{volume}{63}},
  \bibinfo{pages}{205104} (\bibinfo{year}{2001}).

\bibitem[{\citenamefont{Strong and
  Millis}(1994)}]{Strong94_Competition_between_Heisenberg_and_Kondo_in_1D}
\bibinfo{author}{\bibfnamefont{S.~P.} \bibnamefont{Strong}} \bibnamefont{and}
  \bibinfo{author}{\bibfnamefont{A.~J.} \bibnamefont{Millis}},
  \bibinfo{journal}{Phys. Rev. B} \textbf{\bibinfo{volume}{50}},
  \bibinfo{pages}{9911} (\bibinfo{year}{1994}).

\bibitem[{\citenamefont{Zachar et~al.}(1996)\citenamefont{Zachar, Kivelson, and
  Emery}}]{zachar_kondo_chain_toulouse}
\bibinfo{author}{\bibfnamefont{O.}~\bibnamefont{Zachar}},
  \bibinfo{author}{\bibfnamefont{S.~A.} \bibnamefont{Kivelson}},
  \bibnamefont{and} \bibinfo{author}{\bibfnamefont{V.~J.} \bibnamefont{Emery}},
  \bibinfo{journal}{Phys. Rev. Lett.} \textbf{\bibinfo{volume}{77}},
  \bibinfo{pages}{1342} (\bibinfo{year}{1996}).

\bibitem[{\citenamefont{Sikkema et~al.}(1997)\citenamefont{Sikkema, Affleck,
  and White}}]{Sikkema97_Spin_gap_in_a_doped_Kondo_chain}
\bibinfo{author}{\bibfnamefont{A.~E.} \bibnamefont{Sikkema}},
  \bibinfo{author}{\bibfnamefont{I.}~\bibnamefont{Affleck}}, \bibnamefont{and}
  \bibinfo{author}{\bibfnamefont{S.~R.} \bibnamefont{White}},
  \bibinfo{journal}{Phys. Rev. Lett.} \textbf{\bibinfo{volume}{79}},
  \bibinfo{pages}{929} (\bibinfo{year}{1997}).

\bibitem[{\citenamefont{Tsunetsugu et~al.}(1997)\citenamefont{Tsunetsugu,
  Sigrist, and Ueda}}]{tsunetsugu_kondo_1d}
\bibinfo{author}{\bibfnamefont{H.}~\bibnamefont{Tsunetsugu}},
  \bibinfo{author}{\bibfnamefont{M.}~\bibnamefont{Sigrist}}, \bibnamefont{and}
  \bibinfo{author}{\bibfnamefont{K.}~\bibnamefont{Ueda}},
  \bibinfo{journal}{Rev. Mod. Phys.} \textbf{\bibinfo{volume}{69}},
  \bibinfo{pages}{809} (\bibinfo{year}{1997}).

\bibitem[{\citenamefont{Novais et~al.}(2002)\citenamefont{Novais, Miranda,
  Castro~Neto, and
  Cabrera}}]{Novais02_Phase_diagram_of_the_anisotropic_Kondo_chain}
\bibinfo{author}{\bibfnamefont{E.}~\bibnamefont{Novais}},
  \bibinfo{author}{\bibfnamefont{E.}~\bibnamefont{Miranda}},
  \bibinfo{author}{\bibfnamefont{A.~H.} \bibnamefont{Castro~Neto}},
  \bibnamefont{and} \bibinfo{author}{\bibfnamefont{G.~G.}
  \bibnamefont{Cabrera}}, \bibinfo{journal}{Phys. Rev. Lett.}
  \textbf{\bibinfo{volume}{88}}, \bibinfo{pages}{217201}
  (\bibinfo{year}{2002}).

\bibitem[{\citenamefont{Braunecker et~al.}(2009)\citenamefont{Braunecker,
  Simon, and Loss}}]{Braunecker09_Nuclear_magnetism_in_C13_nanotubes}
\bibinfo{author}{\bibfnamefont{B.}~\bibnamefont{Braunecker}},
  \bibinfo{author}{\bibfnamefont{P.}~\bibnamefont{Simon}}, \bibnamefont{and}
  \bibinfo{author}{\bibfnamefont{D.}~\bibnamefont{Loss}},
  \bibinfo{journal}{Phys. Rev. Lett.} \textbf{\bibinfo{volume}{102}},
  \bibinfo{pages}{116403} (\bibinfo{year}{2009}).

\bibitem[{\citenamefont{Nozi\`eres}(1974)}]{nozieres74}
\bibinfo{author}{\bibfnamefont{P.}~\bibnamefont{Nozi\`eres}},
  \bibinfo{journal}{J. Low Temp. Phys.} \textbf{\bibinfo{volume}{17}},
  \bibinfo{pages}{31} (\bibinfo{year}{1974}).

\bibitem[{\citenamefont{Takahashi}(1973)}]{takahashi73}
\bibinfo{author}{\bibfnamefont{M.}~\bibnamefont{Takahashi}},
  \bibinfo{journal}{Prog. Theor. Phys.} \textbf{\bibinfo{volume}{50}},
  \bibinfo{pages}{1519} (\bibinfo{year}{1973}).

\bibitem[{\citenamefont{Luther and Peschel}(1975)}]{luther_chaine_xxz}
\bibinfo{author}{\bibfnamefont{A.}~\bibnamefont{Luther}} \bibnamefont{and}
  \bibinfo{author}{\bibfnamefont{I.}~\bibnamefont{Peschel}},
  \bibinfo{journal}{Phys. Rev. B} \textbf{\bibinfo{volume}{12}},
  \bibinfo{pages}{3908} (\bibinfo{year}{1975}).

\bibitem[{\citenamefont{Lobos et~al.}()\citenamefont{Lobos, Cazalilla, and
  Chudzinski}}]{Lobos12_unpublished_Kondo_Heisenberg}
\bibinfo{author}{\bibfnamefont{A.~M.} \bibnamefont{Lobos}},
  \bibinfo{author}{\bibfnamefont{M.~A.} \bibnamefont{Cazalilla}},
  \bibnamefont{and}
  \bibinfo{author}{\bibfnamefont{P.}~\bibnamefont{Chudzinski}},
  \bibinfo{note}{in preparation}.

\bibitem[{\citenamefont{Gogolin
  et~al.}(1999{\natexlab{a}})\citenamefont{Gogolin, Nersesyan, and
  Tsvelik}}]{gogolin_book}
\bibinfo{author}{\bibfnamefont{A.~O.} \bibnamefont{Gogolin}},
  \bibinfo{author}{\bibfnamefont{A.~A.} \bibnamefont{Nersesyan}},
  \bibnamefont{and} \bibinfo{author}{\bibfnamefont{A.~M.}
  \bibnamefont{Tsvelik}}, \emph{\bibinfo{title}{Bosonization and Strongly
  Correlated Systems}} (\bibinfo{publisher}{Cambridge University Press},
  \bibinfo{address}{Cambridge}, \bibinfo{year}{1999}{\natexlab{a}}).

\bibitem[{\citenamefont{Andreani and
  Beck}(1993)}]{Andreani93_Two-impurity_Anderson_model_variational}
\bibinfo{author}{\bibfnamefont{L.~C.} \bibnamefont{Andreani}} \bibnamefont{and}
  \bibinfo{author}{\bibfnamefont{H.}~\bibnamefont{Beck}},
  \bibinfo{journal}{Phys. Rev. B} \textbf{\bibinfo{volume}{48}},
  \bibinfo{pages}{7322} (\bibinfo{year}{1993}).

\bibitem[{\citenamefont{Barzykin and Affleck}(2000)}]{Barzykin00_Kondo_cloud}
\bibinfo{author}{\bibfnamefont{V.}~\bibnamefont{Barzykin}} \bibnamefont{and}
  \bibinfo{author}{\bibfnamefont{I.}~\bibnamefont{Affleck}},
  \bibinfo{journal}{Phys. Rev. B} \textbf{\bibinfo{volume}{61}},
  \bibinfo{pages}{6170} (\bibinfo{year}{2000}).

\bibitem[{\citenamefont{Simonin}(2007)}]{Simonin07_Kondo_cloud}
\bibinfo{author}{\bibfnamefont{J.}~\bibnamefont{Simonin}},
  \bibinfo{journal}{cond-mat/0708.3604}  (\bibinfo{year}{2007}).

\bibitem[{\citenamefont{Nozi{\`e}res}(1985)}]{nozieres_exhaustion}
\bibinfo{author}{\bibfnamefont{P.}~\bibnamefont{Nozi{\`e}res}},
  \bibinfo{journal}{Ann. Phys. (Paris)} \textbf{\bibinfo{volume}{10}},
  \bibinfo{pages}{19} (\bibinfo{year}{1985}).

\bibitem[{\citenamefont{Simonin}(2006)}]{Simonin06_Two_Anderson_impurities_in_the_Kondo_limit}
\bibinfo{author}{\bibfnamefont{J.}~\bibnamefont{Simonin}},
  \bibinfo{journal}{Phys. Rev. B} \textbf{\bibinfo{volume}{73}},
  \bibinfo{pages}{155102} (\bibinfo{year}{2006}).

\bibitem[{\citenamefont{Schlottmann}(1978)}]{schlottmann_transformation_kondo}
\bibinfo{author}{\bibfnamefont{P.}~\bibnamefont{Schlottmann}},
  \bibinfo{journal}{J. Phys. (Paris)} \textbf{\bibinfo{volume}{C6}},
  \bibinfo{pages}{1486} (\bibinfo{year}{1978}).

\bibitem[{\citenamefont{Emery and
  Kivelson}(1994)}]{emery_kivelson_kondo_review}
\bibinfo{author}{\bibfnamefont{V.~J.} \bibnamefont{Emery}} \bibnamefont{and}
  \bibinfo{author}{\bibfnamefont{S.~A.} \bibnamefont{Kivelson}}, in
  \emph{\bibinfo{booktitle}{Fundamental Problems in Statistical Mechanics VII:
  Proceedings of the 1993 Altenberg Summer School}}, edited by
  \bibinfo{editor}{\bibfnamefont{H.}~\bibnamefont{van Beijeren}}
  \bibnamefont{and} \bibinfo{editor}{\bibfnamefont{M.~E.} \bibnamefont{Ernst}}
  (\bibinfo{publisher}{North Holland}, \bibinfo{address}{Amsterdam},
  \bibinfo{year}{1994}).

\bibitem[{\citenamefont{Kotliar and
  Si}(1996)}]{Kotliar96_Toulouse_points_in_the_generalized_Anderson_model}
\bibinfo{author}{\bibfnamefont{G.}~\bibnamefont{Kotliar}} \bibnamefont{and}
  \bibinfo{author}{\bibfnamefont{Q.}~\bibnamefont{Si}}, \bibinfo{journal}{Phys.
  Rev. B} \textbf{\bibinfo{volume}{53}}, \bibinfo{pages}{12373}
  (\bibinfo{year}{1996}).

\bibitem[{\citenamefont{Guinea et~al.}(1985)\citenamefont{Guinea, Hakim, and
  Muramatsu}}]{Guinea85_Bosonization_of_a_two_level_system}
\bibinfo{author}{\bibfnamefont{F.}~\bibnamefont{Guinea}},
  \bibinfo{author}{\bibfnamefont{V.}~\bibnamefont{Hakim}}, \bibnamefont{and}
  \bibinfo{author}{\bibfnamefont{A.}~\bibnamefont{Muramatsu}},
  \bibinfo{journal}{Phys. Rev. B} pp. \bibinfo{pages}{4410--4418}
  (\bibinfo{year}{1985}).

\bibitem[{\citenamefont{Leggett et~al.}(1987)\citenamefont{Leggett,
  Chakravarty, Dorsey, Fisher, Garg, and Zwerger}}]{leggett_two_state}
\bibinfo{author}{\bibfnamefont{A.~J.} \bibnamefont{Leggett}},
  \bibinfo{author}{\bibfnamefont{S.}~\bibnamefont{Chakravarty}},
  \bibinfo{author}{\bibfnamefont{A.~T.} \bibnamefont{Dorsey}},
  \bibinfo{author}{\bibfnamefont{M.~P.~A.} \bibnamefont{Fisher}},
  \bibinfo{author}{\bibfnamefont{A.}~\bibnamefont{Garg}}, \bibnamefont{and}
  \bibinfo{author}{\bibfnamefont{W.}~\bibnamefont{Zwerger}},
  \bibinfo{journal}{Rev. Mod. Phys.} \textbf{\bibinfo{volume}{59}},
  \bibinfo{pages}{1} (\bibinfo{year}{1987}).

\bibitem[{\citenamefont{Weiss}(1999)}]{Weiss99_Quantum_Dissipative_Systems}
\bibinfo{author}{\bibfnamefont{U.}~\bibnamefont{Weiss}},
  \emph{\bibinfo{title}{Quantum Dissipative Systems (2nd edition)}},
  vol.~\bibinfo{volume}{10} (\bibinfo{publisher}{World Scientific Publishing
  Co. Pte. Ltd., Singapore}, \bibinfo{year}{1999}).

\bibitem[{\citenamefont{Guinea et~al.}(1997)\citenamefont{Guinea, Bascones, and
  Calderon}}]{Guinea97_Quantum_dissipative_systems_review}
\bibinfo{author}{\bibfnamefont{F.}~\bibnamefont{Guinea}},
  \bibinfo{author}{\bibfnamefont{E.}~\bibnamefont{Bascones}}, \bibnamefont{and}
  \bibinfo{author}{\bibfnamefont{M.~J.} \bibnamefont{Calderon}},
  \bibinfo{journal}{AIP Conf. Proc.} \textbf{\bibinfo{volume}{438}},
  \bibinfo{pages}{1} (\bibinfo{year}{1997}).

\bibitem[{\citenamefont{Le~Hur}(2008)}]{LeHur08_Dissipative_Two-level_systems_review}
\bibinfo{author}{\bibfnamefont{K.}~\bibnamefont{Le~Hur}},
  \bibinfo{journal}{Annals of Physics} \textbf{\bibinfo{volume}{323}},
  \bibinfo{pages}{2208} (\bibinfo{year}{2008}).

\bibitem[{\citenamefont{Gogolin
  et~al.}(1999{\natexlab{b}})\citenamefont{Gogolin, Nersesyan, and
  Tsvelik}}]{gogolin_1dbook}
\bibinfo{author}{\bibfnamefont{A.~O.} \bibnamefont{Gogolin}},
  \bibinfo{author}{\bibfnamefont{A.~A.} \bibnamefont{Nersesyan}},
  \bibnamefont{and} \bibinfo{author}{\bibfnamefont{A.~M.}
  \bibnamefont{Tsvelik}}, \emph{\bibinfo{title}{Bosonization and Strongly
  Correlated Systems}} (\bibinfo{publisher}{Cambridge University Press},
  \bibinfo{address}{Cambridge}, \bibinfo{year}{1999}{\natexlab{b}}).

\bibitem[{Lob({\natexlab{a}})}]{Lobos12_comment_1D_Kondo_cloud}
\bibinfo{note}{This is in contrast to spins interacting with a 1D metal, where
  the single-impurity limit is only reached at much larger distances
  $d\gg\xi_{K}$ , with $\xi_{K}=\hbar v_{F}/k_{B}T_{K}$ the Kondo correlation
  length and $T_{K}$ the Kondo temperature.}

\bibitem[{\citenamefont{Drewes et~al.}(2003)\citenamefont{Drewes, Arovas, and
  Renn}}]{drewes03_dissipative_jj}
\bibinfo{author}{\bibfnamefont{S.}~\bibnamefont{Drewes}},
  \bibinfo{author}{\bibfnamefont{D.~P.} \bibnamefont{Arovas}},
  \bibnamefont{and} \bibinfo{author}{\bibfnamefont{S.}~\bibnamefont{Renn}},
  \bibinfo{journal}{Phys. Rev. B} \textbf{\bibinfo{volume}{68}},
  \bibinfo{pages}{165345} (\bibinfo{year}{2003}).

\bibitem[{\citenamefont{Renn}(1995)}]{Renn97_condmat_dissipative_quantum_rotors}
\bibinfo{author}{\bibfnamefont{S.~R.} \bibnamefont{Renn}},
  \bibinfo{journal}{cond-mat/9708194}  (\bibinfo{year}{1995}).

\bibitem[{\citenamefont{Werner et~al.}(2005)\citenamefont{Werner, Troyer, and
  Sachdev}}]{Werner05_Quantum_Spin_Chains_with_site_dissipation}
\bibinfo{author}{\bibfnamefont{P.}~\bibnamefont{Werner}},
  \bibinfo{author}{\bibfnamefont{M.}~\bibnamefont{Troyer}}, \bibnamefont{and}
  \bibinfo{author}{\bibfnamefont{S.}~\bibnamefont{Sachdev}},
  \bibinfo{journal}{J. Phys. Soc. Jpn.} \textbf{\bibinfo{volume}{74}},
  \bibinfo{pages}{67} (\bibinfo{year}{2005}).

\bibitem[{\citenamefont{Feigel'man and
  Larkin}(1998)}]{feigelman98_smt_in_2d_proximity_array}
\bibinfo{author}{\bibfnamefont{M.~V.} \bibnamefont{Feigel'man}}
  \bibnamefont{and} \bibinfo{author}{\bibfnamefont{A.~I.}
  \bibnamefont{Larkin}}, \bibinfo{journal}{Chem. Phys.}
  \textbf{\bibinfo{volume}{235}}, \bibinfo{pages}{107} (\bibinfo{year}{1998}).

\bibitem[{\citenamefont{Sachdev}(2000)}]{sachdev_book}
\bibinfo{author}{\bibfnamefont{S.}~\bibnamefont{Sachdev}},
  \emph{\bibinfo{title}{Quantum Phase Transitions}}
  (\bibinfo{publisher}{Cambridge University Press},
  \bibinfo{address}{Cambridge, UK}, \bibinfo{year}{2000}).

\bibitem[{\citenamefont{Lutchyn et~al.}(2008)\citenamefont{Lutchyn, Galitski,
  Refael, and Das~Sarma}}]{Lutchyn08_Dissipative_QPT_in_SC-graphene_systems}
\bibinfo{author}{\bibfnamefont{R.~M.} \bibnamefont{Lutchyn}},
  \bibinfo{author}{\bibfnamefont{V.}~\bibnamefont{Galitski}},
  \bibinfo{author}{\bibfnamefont{G.}~\bibnamefont{Refael}}, \bibnamefont{and}
  \bibinfo{author}{\bibfnamefont{S.}~\bibnamefont{Das~Sarma}},
  \bibinfo{journal}{Phys. Rev. Lett.} \textbf{\bibinfo{volume}{101}},
  \bibinfo{pages}{106402} (\bibinfo{year}{2008}).

\bibitem[{\citenamefont{Mahan}(2000)}]{mahan2000}
\bibinfo{author}{\bibfnamefont{G.~D.} \bibnamefont{Mahan}},
  \emph{\bibinfo{title}{Many-Particle Physics}}, Physics of Solids and Liquids
  (\bibinfo{publisher}{Kluwer Academic/Plenum Publishers},
  \bibinfo{address}{New York}, \bibinfo{year}{2000}), \bibinfo{edition}{3rd}
  ed.

\bibitem[{\citenamefont{Pankov et~al.}(2004)\citenamefont{Pankov, Florens,
  Georges, Kotliar, and
  Sachdev}}]{Pankov04_NFL_behavior_and_2D_AFM_fluctuations}
\bibinfo{author}{\bibfnamefont{S.}~\bibnamefont{Pankov}},
  \bibinfo{author}{\bibfnamefont{S.}~\bibnamefont{Florens}},
  \bibinfo{author}{\bibfnamefont{A.}~\bibnamefont{Georges}},
  \bibinfo{author}{\bibfnamefont{G.}~\bibnamefont{Kotliar}}, \bibnamefont{and}
  \bibinfo{author}{\bibfnamefont{S.}~\bibnamefont{Sachdev}},
  \bibinfo{journal}{Phys. Rev. B} \textbf{\bibinfo{volume}{69}},
  \bibinfo{pages}{054426} (\bibinfo{year}{2004}).

\bibitem[{\citenamefont{Hertz}(1976)}]{Hertz76_Quantum_Critical_Phenomena}
\bibinfo{author}{\bibfnamefont{J.~A.} \bibnamefont{Hertz}},
  \bibinfo{journal}{Phys. Rev. B} \textbf{\bibinfo{volume}{14}},
  \bibinfo{pages}{1165} (\bibinfo{year}{1976}).

\bibitem[{\citenamefont{Moriya and
  Kawabata}(1973)}]{Moriya73_Critical_phenomena}
\bibinfo{author}{\bibfnamefont{T.}~\bibnamefont{Moriya}} \bibnamefont{and}
  \bibinfo{author}{\bibfnamefont{J.}~\bibnamefont{Kawabata}},
  \bibinfo{journal}{J. Phys. Soc. Jpn.} \textbf{\bibinfo{volume}{34}},
  \bibinfo{pages}{639} (\bibinfo{year}{1973}).

\bibitem[{\citenamefont{Millis}(1993)}]{Millis93_Quantum_critical_points_in_Fermi_systems}
\bibinfo{author}{\bibfnamefont{A.~J.} \bibnamefont{Millis}},
  \bibinfo{journal}{Phys. Rev. B} \textbf{\bibinfo{volume}{48}},
  \bibinfo{pages}{7183} (\bibinfo{year}{1993}).

\bibitem[{\citenamefont{Mermin and Wagner}(1967)}]{mermin_wagner_theorem}
\bibinfo{author}{\bibfnamefont{N.~D.} \bibnamefont{Mermin}} \bibnamefont{and}
  \bibinfo{author}{\bibfnamefont{H.}~\bibnamefont{Wagner}},
  \bibinfo{journal}{Phys. Rev. Lett.} \textbf{\bibinfo{volume}{17}},
  \bibinfo{pages}{1133} (\bibinfo{year}{1967}).

\bibitem[{\citenamefont{Feynman}(1972)}]{feynman_statmech}
\bibinfo{author}{\bibfnamefont{R.~P.} \bibnamefont{Feynman}},
  \emph{\bibinfo{title}{Statistical Mechanics}} (\bibinfo{publisher}{Benjamin},
  \bibinfo{address}{Reading, MA}, \bibinfo{year}{1972}).

\bibitem[{\citenamefont{Hirjibehedin et~al.}(2007)\citenamefont{Hirjibehedin,
  Lin, Otte, Ternes, Lutz, Jones, and
  Heinrich}}]{Hirjibehedin07_Magnetic_anisotropy_in_surfaces}
\bibinfo{author}{\bibfnamefont{C.~F.} \bibnamefont{Hirjibehedin}},
  \bibinfo{author}{\bibfnamefont{C.-Y.} \bibnamefont{Lin}},
  \bibinfo{author}{\bibfnamefont{A.~F.} \bibnamefont{Otte}},
  \bibinfo{author}{\bibfnamefont{M.}~\bibnamefont{Ternes}},
  \bibinfo{author}{\bibfnamefont{C.~P.} \bibnamefont{Lutz}},
  \bibinfo{author}{\bibfnamefont{B.~A.} \bibnamefont{Jones}}, \bibnamefont{and}
  \bibinfo{author}{\bibfnamefont{A.~J.} \bibnamefont{Heinrich}},
  \bibinfo{journal}{Science} \textbf{\bibinfo{volume}{317}},
  \bibinfo{pages}{1199} (\bibinfo{year}{2007}).

\bibitem[{\citenamefont{Barral et~al.}(2010)\citenamefont{Barral, Roura-Bas,
  Llois, and Aligia}}]{Barral10_Anisotropic_Kondo}
\bibinfo{author}{\bibfnamefont{M.~A.} \bibnamefont{Barral}},
  \bibinfo{author}{\bibfnamefont{P.}~\bibnamefont{Roura-Bas}},
  \bibinfo{author}{\bibfnamefont{A.~M.} \bibnamefont{Llois}}, \bibnamefont{and}
  \bibinfo{author}{\bibfnamefont{A.~A.} \bibnamefont{Aligia}},
  \bibinfo{journal}{Phys. Rev. B} \textbf{\bibinfo{volume}{82}},
  \bibinfo{pages}{125438} (\bibinfo{year}{2010}).

\bibitem[{\citenamefont{Akkermans and Montambaux}(2007)}]{akkermans}
\bibinfo{author}{\bibfnamefont{E.}~\bibnamefont{Akkermans}} \bibnamefont{and}
  \bibinfo{author}{\bibfnamefont{G.}~\bibnamefont{Montambaux}},
  \emph{\bibinfo{title}{Mesoscopic Physics of Electrons and Photons}}
  (\bibinfo{publisher}{Cambridge University Press},
  \bibinfo{address}{Cambridge}, \bibinfo{year}{2007}).

\bibitem[{Lob({\natexlab{b}})}]{Lobos12_comment_sign_JH}
\bibinfo{note}{Strictly speaking, for the analogy to be correct, $J_H^\perp$
  should have the opposite sign. This can be done introducing the trivial
  change of variables $\varphi_i =\varphi_i^\prime +i\pi $.}

\end{thebibliography}

\end{document}